\documentclass{emulateapj}

\newcommand{\HI}{\ion{H}{1}~}

\newcommand{\kms}{km~s$^{-1}$ }
\newcommand{\Lya}{Ly$\alpha$~}

\usepackage{lscape} 
\usepackage{color} 
\usepackage{graphics,graphicx}

\slugcomment{Accepted for publication in ApJ}

\shorttitle{Connecting the CGM to the ISM in galaxies}
\shortauthors{Borthakur et al.}

\begin{document}

\title{The Connection Between the Circumgalactic Medium and the Interstellar Medium of Galaxies: Results from the COS-GASS Survey}

\author{Sanchayeeta Borthakur}
\affil{Department of Physics \& Astronomy, Johns Hopkins University, Baltimore, MD, 21218, USA and The Astronomy Department, University of California, Berkeley, CA USA }
\email{sanch@pha.jhu.edu}

\author{Timothy Heckman}
\affil{Department of Physics \& Astronomy, Johns Hopkins University, Baltimore, MD, 21218, USA }

\author{Jason Tumlinson, Rongmon Bordoloi, Christopher Thom}
\affil{Space Telescope Science Institute, Baltimore, MD 21218, USA}

\author{Barbara Catinella}
\affil{Centre for Astrophysics \& Supercomputing, Swinburne University of Technology, Hawthorn, VIC 3122, Australia}

\author{David Schiminovich}
\affil{Department of Astronomy, Columbia University, New York, NY 10027, USA}

\author{Romeel Dav\scriptsize{$\rm \acute{E}$}}
\affil{ University of the Western Cape, Bellville, Cape Town 7535, \\
South Africa, South African Astronomical Observatories, Observatory, Cape Town 7925, \\
South Africa, African Institute for Mathematical Sciences, Muizenberg, Cape Town 7945, South Africa }

\author{Guinevere Kauffmann}
\affil{Max-Planck Institut f$\rm\ddot{u}$r Astrophysik, D-85741 Garching, Germany }

\author{Sean M. Moran}
\affil{Harvard-Smithsonian Center for Astrophysics, 60 Garden Street, Cambridge, MA 02138}

\author{Amelie Saintonge}
\affil{Department of Physics and Astronomy, University College London, Gower Place, London WC1E 6BT, UK}

\begin{abstract}

We present a study exploring the nature and properties of the Circum-Galactic Medium (CGM) and its connection to the atomic gas content in the interstellar medium (ISM) of galaxies as traced by the HI 21~cm line. Our sample includes 45 low-z (0.026-0.049) galaxies from the GALEX Arecibo SDSS Survey. Their CGM was probed via absorption in the spectra of background Quasi-Stellar Objects at impact parameters of 63 to 231~kpc. The spectra were obtained with the Cosmic Origins Spectrograph aboard the Hubble Space Telescope. We detected neutral hydrogen (\Lya absorption-lines) in the CGM of 92\% of the galaxies. We find the radial profile of the CGM as traced by the \Lya equivalent width can be fit as an exponential with a scale length of roughly the virial radius of the dark matter halo. We found no correlation between the orientation of sightline relative to the galaxy major axis and the \Lya equivalent width. The velocity spread of the circumgalactic gas is consistent with that seen in the atomic gas in the interstellar medium.  We find a strong correlation (99.8\% confidence) between the gas fraction (M(HI)/M$_{\star}$) and the impact-parameter-corrected \Lya equivalent width. This is stronger than the analogous correlation between corrected \Lya equivalent width and SFR/M$_{\star}$ (97.5\% confidence). These results imply a physical connection between the \HI disk and the CGM, which is on scales an order-of-magnitude larger. This is consistent with the picture in which the \HI disk is nourished by accretion of gas from the CGM.

\end{abstract}


\section{INTRODUCTION\label{intro}}

In the standard paradigm, galaxies grow primarily through the accretion of gas that flows from the Inter-Galactic Medium (IGM). Much of this accreted gas is ultimately turned into new stars. In turn, the massive stars release matter and energy that can affect not only the galaxy itself, but also the IGM. These flows in and out of galaxies pass through the Circum-Galactic Medium (CGM), a region extending out from the galaxy itself to roughly the virial radius of the surrounding dark matter halo. The CGM also represents a significant reservoir of baryons \citep{werk14,s13}.
Thus, understanding the properties of the CGM is critical to understanding how galaxies and the IGM co-evolve.

The low densities in the CGM imply that it is very difficult to study in emission. Our ability to study the CGM in the present-day universe has been dramatically improved with the installation of the Cosmic Origins Spectrograph (COS) on the Hubble Space Telescope (HST). This has enabled us to use the rich suite of absorption-lines due to resonance transitions of many important atoms and ions that are present in the vacuum ultraviolet. Indeed, over the last few years, studies such as the COS-Halos, COS-Dwarfs, and others \citep{tumlinson11a, tripp11, prochaska11, borthakur13, tumlinson13, s13, werk14, bordoloi14, liang14, johnson15}
 have established the connection between the CGM and many of the key properties of galaxies such as star-formation rate (SFR), stellar mass, color etc.  

However, so far we have yet to explore the connection between the interstellar medium (ISM) and the CGM. This is important because in the simplest picture we would expect that incoming gas from the CGM would be largely ionized  \citep[e.g.][]{werk14} and upon accretion would then have to pass through an HI phase and then a molecular phase before being converted into stars. Thus, the first step in assessing the role of accretion from the CGM in the evolution of galaxies would be to look at the relationship between the properties of the CGM and of the atomic gas in the galaxy disk.

 \begin{figure*}
\begin{center}
\includegraphics[trim = 0mm 10mm 0mm 0mm, clip,scale=0.75,angle=-0]{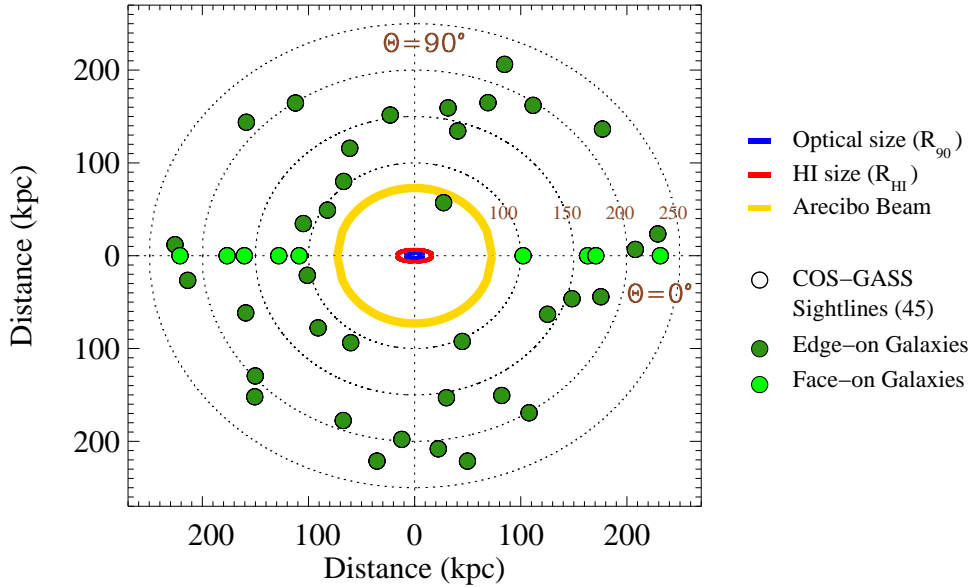}  
\end{center}
\caption{Overview of the COS-GASS sample. The figure shows a representative galaxy from the COS-GASS sample at the center of the plot and QSO sightlines shown as filled circles. The sample consists of 45 QSO-galaxy pairs probing the CGM in galaxies from the GASS project. The stellar masses of the sample galaxies range between $\rm 10^{10.1-11.1}~M_{\odot}$. The median optical (R$\rm_{90}$) and \textsc{Hi} radii (R$\rm_{HI}$) of the galaxies in the sample are shown as blue and red ellipses respectively at the center of the plot. The physical size corresponding to the FWHM of the Arecibo beam at 21~cm at the average redshift for our sample ($\overline{z}=0.037$) is shown in yellow. The QSO sightlines cover a range of impact parameters from 63-231~kpc. The position of the QSO sightlines with respect to the foreground galaxy are shown as filled green circles. The orientations of sightlines with respect to the disks are estimated using position angle of the galaxy major axis from SDSS photometry. We consider galaxies with de Vaucouleurs a/b parameter greater than 0.7 as having a face-on orientation. These galaxies were assigned an orientation parameter of zero degrees and are identified in this figure with a lighter shade of green.}  
 \label{fig-sample} 
\end{figure*}

Of particular interest in this context is a better understanding of the processes that produce a clear bimodality between the population of lower mass galaxies undergoing significant star-formation and higher mass quiescent galaxies. First discovered at low-redshift 
\citep[e.g.][]{kauffmann03, blanton03, baldry04}, this split into a blue star-forming main sequence and a 'red and dead' population is now known to be in place out to redshifts of two and beyond \citep[e.g.][]{elbaz11, whitaker13}. The galaxy bimodality suggests that galaxies undergo fundamental changes (through accretion and merging) in their gas content and their ability to form new stars after either the galaxy itself or its dark matter halo exceeds a critical mass limit \citep[e.g.][]{lilly13}. This quenching of star formation may be related to a change in the nature of the CGM, from a pathway for rapid ($v_{in} \sim v_{vir}$) accretion of relatively cold gas ($T << T_{virial}$) to slow and inefficient accretion of hot gas ($T \sim T_{virial}$) mediated by radiative cooling \citep[e.g.][]{keres05, keres09, dekel09}.

To help probe the processes responsible for the drastic drop in star formation in massive galaxies and the implied change in the cold gas supply, we have undertaken the GALEX Arecibo SDSS Survey \citep[GASS;][]{catinella10}. The GASS project combines 21~cm \HI spectroscopic data obtained with the Arecibo telescope, optical images and spectroscopy from the Sloan Digital Sky Survey (SDSS), and ultraviolet(UV) imaging with the Galaxy Evolution Explorer (GALEX) to measure the atomic gas and stellar content (star formation rate and history), gas phase metallicity, stellar morphology, and disk rotation speeds of about one thousand galaxies over the stellar mass range from M$_* \sim 10^{10.0 - 11.5}$ M$_{\odot}$ (roughly centered on the transition mass between star-forming and quiescent populations).
We have also obtained molecular gas data from IRAM \citep[COLD GASS;][]{saintonge11} and long-slit optical spectroscopy \citep{moran12} for a portion of the GASS sample.

Our sample of galaxies in this paper was drawn from the GASS parent sample. 
This sample enables us to probe the connection between the ISM and CGM in L$_{\star}$ galaxies in a statistically significant manner for  the first time. In addition, our sample is also complementary to the COS-Halos sample as both the samples probe galaxies of similar stellar masses. While COS-Halos probed the inner CGM, our sample extends the CGM coverage out to the virial radii. 
Detailed descriptions of our sample selection criteria, the COS observations, and data reduction are presented in Section~2.  The results are presented in Section~3. Finally, we summarize our findings  and discuss their implications in Section~4. The cosmological parameters used in this study are $H_0 =70~{\rm km~s}^{-1}~{\rm Mpc}^{-1}$, $\Omega_m = 0.3$, and $\Omega_{\Lambda} = 0.7$.

\section{OBSERVATIONS  \label{sec:observations}}

\subsection{Sample: COS-GASS \label{sec:sample}}

Our sample of galaxies was derived from the GASS Survey. We have observed all the Quasi Stellar Object (QSO) that probe GASS galaxies within a projected separation from the galaxy (impact parameter) of $\le$ 250~kpc. The limiting flux of the background QSO was selected to be GALEX Far-UV AB magnitude, $\rm FUV_{mag} \le 19.0$. This yielded an observed sample of 47 QSO/galaxy pairs. The observations of two of those sightlines were found to be unsuitable for this analysis. The first is a Broad Absorption-Line (BAL) QSO in which it is difficult to perform robust measurements of the absorption features associated with the target galaxy. The second sightline yields a damped \Lya system which subsequent observations show not to be directly associated with the primary GASS target. We will discuss this unique system in Borthakur et al. (in preparation). Hence for the remainder of the paper we concentrate on our final 'COS-GASS' sample of 45 QSO sightlines probing the CGM of low-z galaxies.

Figure~\ref{fig-sample} presents the distribution of our QSO sightlines as a function of impact parameter and orientation with respect to the target galaxy. The sightlines are marked with filled green circles. Those probing face-on galaxies being marked in lighter green. The average optical and \HI disk sizes of the galaxies are shown blue and red respectively. The optical sizes for the galaxies were estimated from the SDSS photometric parameter, $\rm R_{90}$. The physical size of the Arecibo beam (full width at half maximum, FWHM, of 3.3$^\prime$) at the average redshift of 0.037 for our sample is shown as the yellow circle. The Arecibo beam covers the optical extent of every galaxy, but does not extend to the virial radius.

The principal properties of the galaxies are summarized in Table~1. Except where noted below, these parameters are taken from the GASS project  \citep[see][for details]{catinella10,catinella12, catinella13}. GASS also incorporates data products from the SDSS \citep{SDSS_DR8}. For example, position and redshift information are based on SDSS photometry and spectroscopic measurements respectively. The errors in the redshift measurements are of the order of $\rm 10^{-4}$, which corresponds to $\sim$ 30~\kms. The values are presented in Table~1 and have been rounded off to the forth decimal place.
The stellar masses of the galaxies in our sample range from $\rm 10^{10.1-11.1}~M_{\odot}$. 
 We estimated dark matter halo masses using the stellar mass measurements from GASS and the methodology described in \citet{behroozi10}(see Eq.21 and Table~2). The halo masses were estimated for z=0 \citep[see for a detailed discussion][]{shull14} and they range from 11.6-13.3 $\rm M_{\odot}$. Based on these halo masses we estimated the virial radii. Our virial radius estimates are within 0.1 dex of those based on the luminosity based prescription in \citet{prochaska11}. We report the halo masses and their corresponding virial radii in Table~1. The SFR and by extension the specific SFR (sSFR = SFR/M$_{\star}$) for the GASS sample were derived using a combination of GALEX $FUV$ and $NUV$ and SDSS $u,g,r,i,z$ photometry and SDSS spectral-line indices.  \citep[see section 2.2 of ][]{schiminovich10}.  Any value for sSFR $ \rm < 10^{-12}~ yr^{-1}$ can be regarded as a upper limit. We quote an uncertainity of 0.2~dex in sSFR for the galaxies with  sSFR $ \rm > 10^{-12}~ yr^{-1}$. These are based on results shown and discussed in detail by  \citet{schiminovich10}. By including UV fluxes (from GALEX) in addition to H$\alpha$ (from SDSS), the derived star formation rates are sensitive to timescales of the order of 10$\rm ^8$~yrs or larger. These rates define the recent star formation activity that might have impacted the CGM. For example, star-formation driven outflows traveling at 200~\kms would reach our closest sightline at an impact parameter of 63~kpc in about 3$\rm \times 10^8$~yrs and sightlines at impact parameter of 200~kpc in $10^9$~yrs.

Information on the QSO sightlines is presented in Table~2. The azimthual orientations of the QSO sightlines with respect to target galaxies were calculated from the position angle of the optical major axis of the galaxy from SDSS r-band photometry. We consider galaxies with the de Vaucouleurs AB parameter greater than 0.7 to be face-on and assign an orientation parameter of 0$^\circ$ in the analysis presented in the remainder of the paper.

\subsection{\HI masses from single-dish measurements }

Our 21~cm \HI spectroscopic data were primarily derived from the GASS observations obtained with the single-dish Arecibo telescope. The GASS sample also included a few galaxies from the Arecibo Legacy Fast ALFA  (ALFALFA) Survey \citep{haynes11} and  Cornell \HI archive \citep[CHA,][]{springob05} that met the GASS sample selection criteria.  GASS HI parameters were measured by smoothing the \HI profiles followed by a baseline subtraction. 
The edges and peaks of the \HI profiles were interactively identified. Further details on the analysis procedures were described by \citet{springob05} and Catinella, Haynes \& Giovanelli 2007 (AJ 134, 334). The velocities at zero flux were marked interactively as part of the width and flux measurement process, at or near the position where the fits to the edges of the profile cross the baseline. The velocity width were corrected for cosmological redshift and instrumental broadening using procedure described in \citet{catinella10}. These measurements did not include corrections for effects due to inclination or turbulent motion. The measurements of \HI masses and widths from the GASS database are included in Table~1.

For five galaxies in our sample, Arecibo \HI measurements were not available (although they were members of the GASS parent sample). For these we performed similar observations with the Robert C. Byrd Green Bank Telescope (GBT)\footnote{The National Radio Astronomy Observatory is a facility of the National Science Foundation operated under cooperative agreement by Associated Universities, Inc.}. 
The GBT data were obtained under program GBT-14A-377. We used the dual polarization L-band system with two intermediate frequency (IF) modes and nine- level sampling. The IFs were set to yield a channel width of 1.56 kHz (0.33~\kms) using 8192 channels over a total bandwidth of 12.5 MHz. Observations were made in the standard position-switching ON-OFF scheme with 300~s at each of the positions. Data were recorded at 10~s intervals to minimize the effect of RFI. The OFF position was chosen +20$^{\prime}$ offset in Right Ascension from each of the sight lines. During each of the observing sessions, we used one of the three standard flux calibrators 3C 48 (16.5~Jy), 3C 147 (22.5~Jy), and 3C 286 (15.0~Jy) - for pointing and estimating antenna gain. Local pointing corrections (LPCs) were performed using the observing procedure AutoPeak and the corrections were then automatically applied to the data. This should result in a pointing accuracy of 3$^{\prime\prime}$. 
 It is worth noting that the pointing accuracy does not signify the positional accuracy of the source of the \HI flux. The \HI emitting region may be present anywhere within the beam. The flux calibrations were performed by applying antenna gain to data for each session separately. Our calibration error is expected to be no more than 10\% . The data were analyzed using the NRAO software GBTIDL.

\begin{figure*}
\begin{center}
\includegraphics[trim = 0mm 0mm 0mm 0mm, clip,scale=0.45,angle=-0]{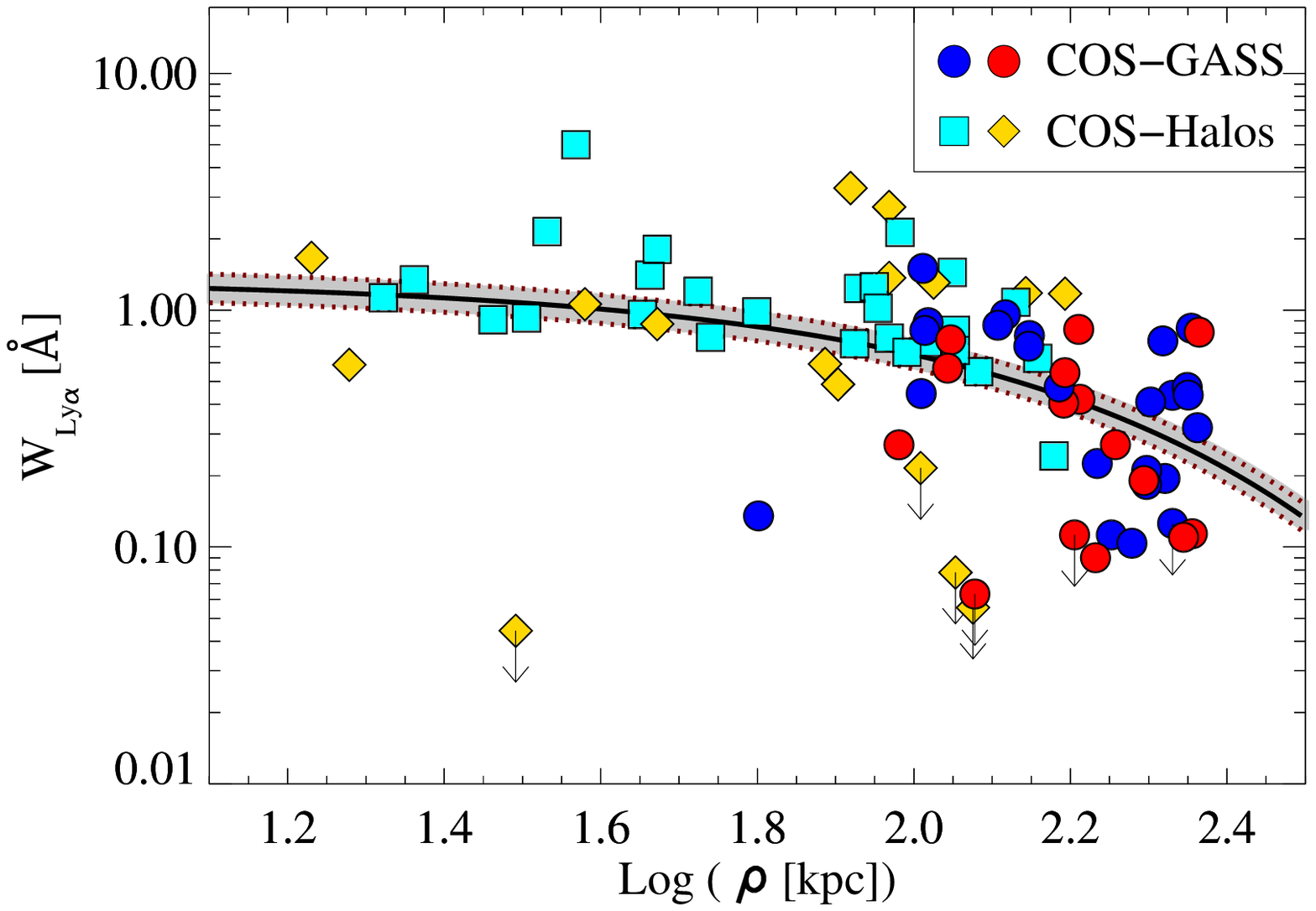}  \\
\includegraphics[trim = 0mm 0mm 0mm 0mm, clip,scale=0.45,angle=-0]{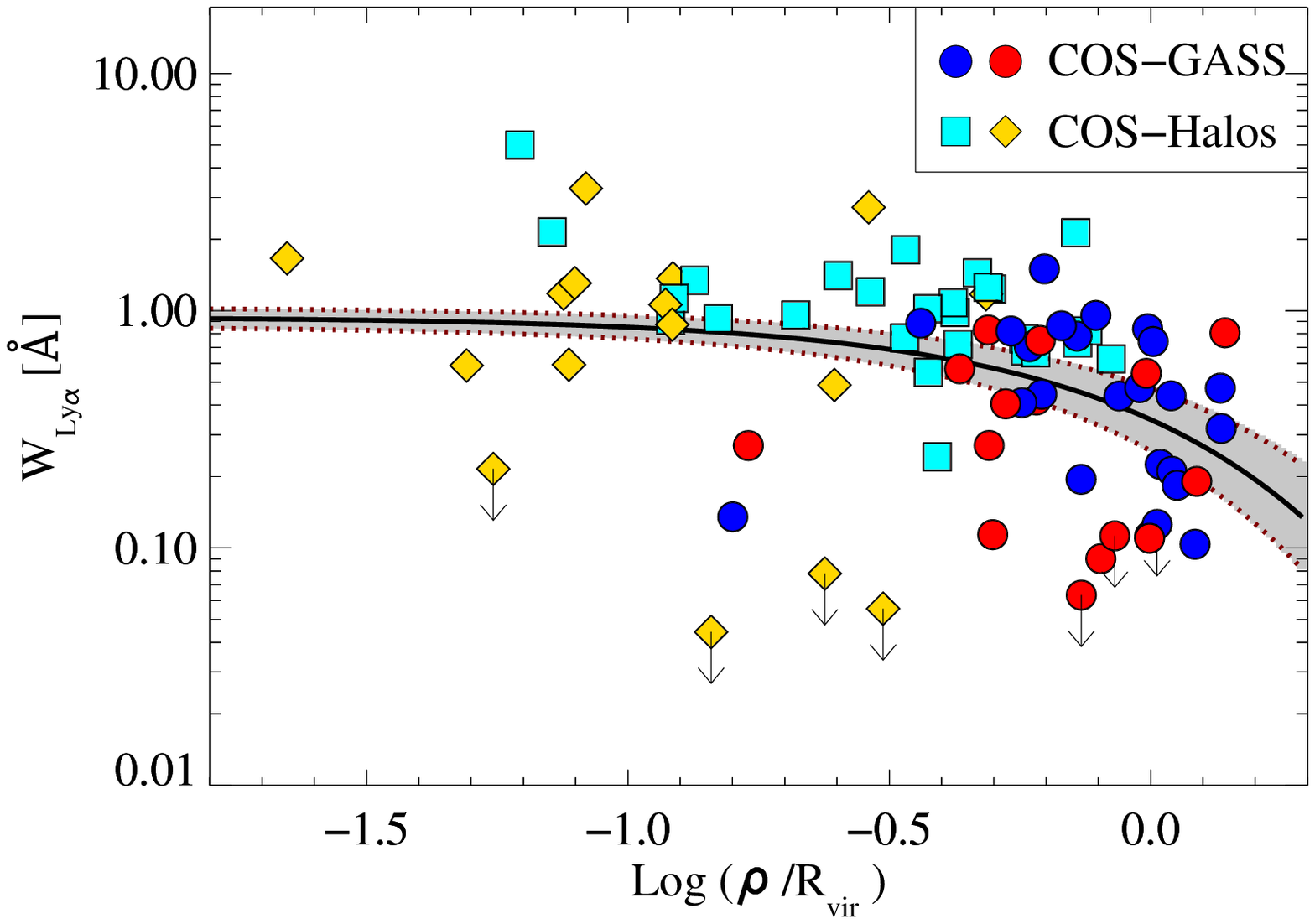} 
\includegraphics[trim = 0mm 0mm 0mm 0mm, clip,scale=0.45,angle=-0]{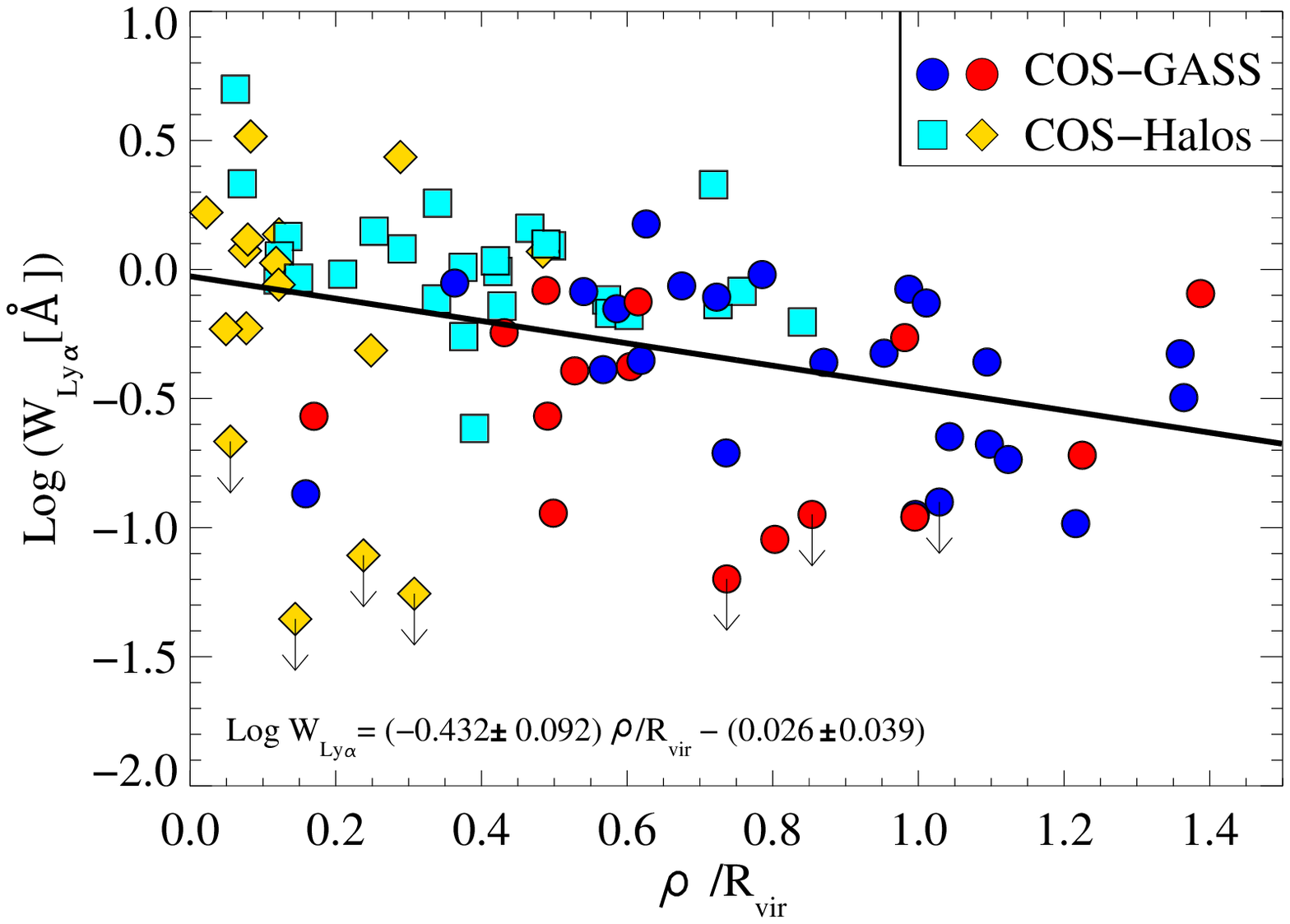}  \\
\end{center}
\caption{Top: Variation of \Lya equivalent widths as a function of impact parameter i.e. $\rm \rho$. The COS-GASS data are shown as circles and the COS-Halos data are shown as diamonds. The galaxies are divided into two bins based on their specific star-formation rates (see text for details). This is indicated by the color of the symbols. Blue and cyan indicate blue star-forming galaxies from the COS-GASS and COS-Halos samples respectively. Red and yellow indicate the red passive galaxies. As shown with the black line, the radial distribution of the \Lya equivalent width can be fit as exponential with an e-folding scale length of 136~kpc. The grey shaded region represents the uncertainty associated with the fit.Bottom: Variation of \Lya equivalent widths as a function of normalized impact parameter i.e. $\rm \rho/R_{vir}$. The panel on the left presents data using a logarithmic radial scale where as the one on the right is a linear radial scale. The best-fit exponential distribution is illustrated with the black solid line. The uncertainty associated with the fit is shown in grey. The best-fit parameters are  labeled at the  bottom of the right panel and imply an e-folding scale length of 1.0~$\rm R_{vir}$.}  
 \label{fig-rho_W} 
\end{figure*}

The GBT observational setup was very similar to that of the GASS program and the data products were measured at the same spectral resolution as the GASS sample. The \HI masses, the velocity widths and the origin of the \HI data are presented in Table~1. The widths represent the full velocity extent of the features i.e. the width of the 21~cm transition at zero flux. The line-widths were estimated visually and the uncertainties associated with these values are typically $\sim$20~\kms. The \HI 21~cm velocity ranges cover the systemic velocity of the galaxy in all cases. We define the systemic velocity as the optical velocity that traces the stars. The stars represent the bulk of the measurable baryons in our galaxies. BY doing so, we also alleviates the problem of identifying the systemic velocity from \HI 21~cm profiles that may be asymmetric.

\subsection{Cosmic Origins Spectrograph Observations}

We carried out ultraviolet spectroscopic observations of the background QSOs with the COS aboard the HST. The data were obtained with the grating G130M of COS, providing a resolution R$= \rm 20,000-24,000$ (FWHM $\sim$ 14~\kms). The data covered observed-frame wavelength of 1140-1470~$\rm \AA$. This included transitions such as \Lya ($\rm \lambda 1215.670~\AA$), \ion{C}{2} ($\rm \lambda 1334~\AA$),   \ion{Si}{2}~($\rm\lambda 1190, 1193, \& 1260~\AA$),   \ion{Si}{3}~($\rm \lambda 1206~\AA$), \ion{Si}{4}~($\rm \lambda\lambda1393, 1402~\AA$), and \ion{O}{1}~($\rm \lambda 1302~\AA$). 

The data were reduced using the standard COS pipeline. All absorption features in the spectra were identified and were exhaustively matched to identify the species and transition. Absorption features associated with the Milky Way, the target galaxies, intervening systems, and the background QSOs were identified. We use a velocity window of $\pm$ 600~\kms  with respect to the systemic velocity (based on SDSS redshifts) to associate absorption features with our target galaxies. In other words, transitions detected within $\pm$ 600~\kms of the systemic velocity of the galaxies are assumed to be associated with the target galaxies. The large velocity window allows us to look for kinematic signatures of possible inflowing or outflowing material. Furthermore, the velocity window is same as used in the other large CGM studies such as COS-Halos \citep{tumlinson13} and COS-Dwarfs \citep{bordoloi14}, thus making our analysis consistent with existing CGM studies.
 
We do miss a few transitions associated with the target galaxies due to blending with other absorption features such as from the Milky Way or intervening absorbers between the QSO and us. In this paper we focus on the \Lya transition probing the hydrogen in the CGM. A paper describing the metal line properties for the galaxies is in preparation.
 The \Lya equivalent width measurements are provided in column 9 of Table~2. 
We fitted Voigt profiles to the \Lya absorption features to estimate the velocity centroid of each of the components contributing to the full profile. Most of the \Lya absorption features (25/36) could be described with a single component. There are 10/36 \Lya absorption profiles that required two components and one that required three. 
 The velocity centroids of all the components (listed in order of the strength of the component) and the entire widths of the \Lya absorption profiles within the rest-frame of the galaxy are also provided in Table~2.

\begin{figure}
\begin{center}
\includegraphics[trim = 0mm 0mm 0mm 0mm, clip,scale=0.55,angle=-0]{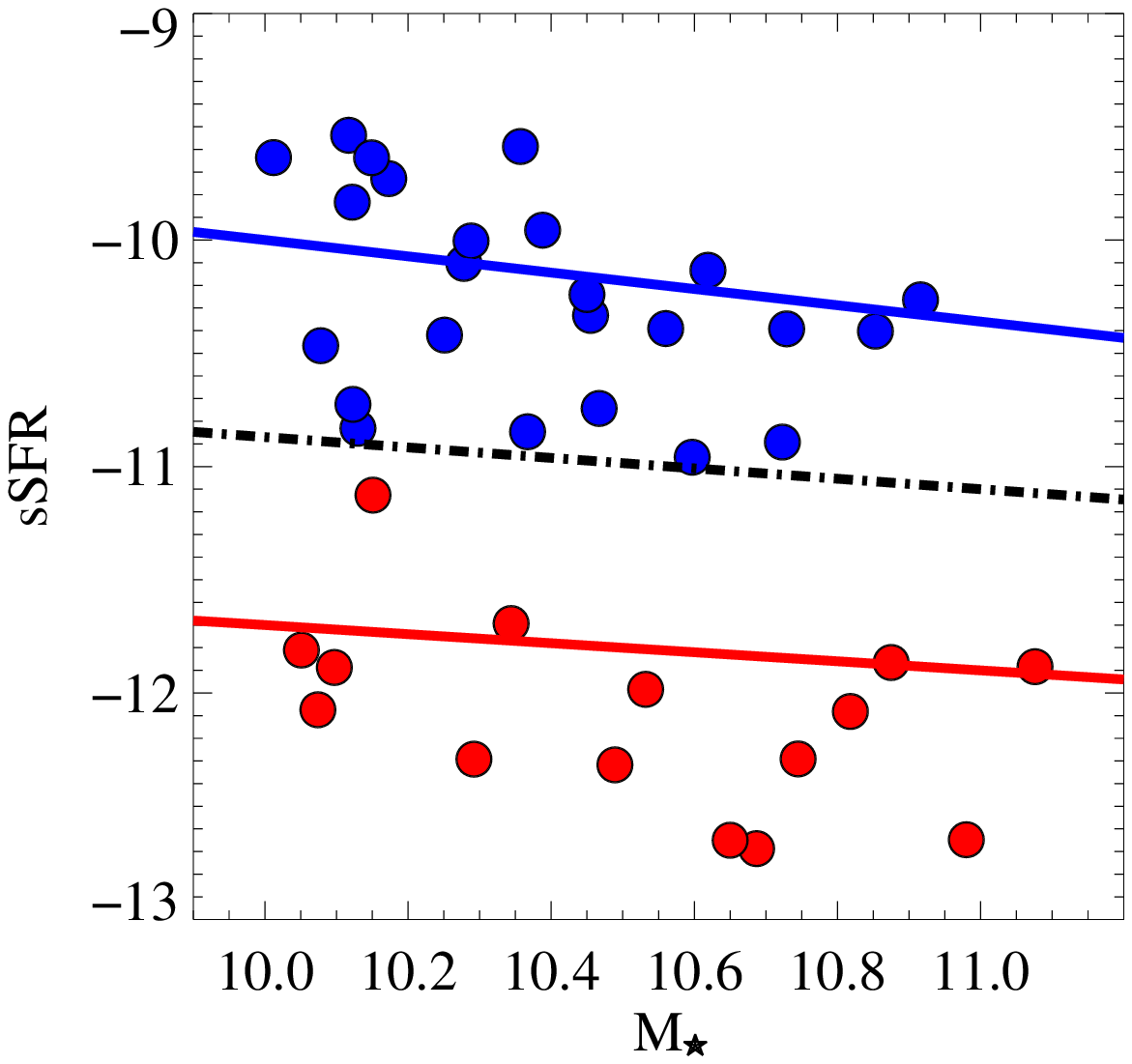}  
\end{center}
\caption{ Galaxy color definition for the COS-GASS sample. The COS-GASS sample on the parameter space of stellar mass ($\rm M_{\star}$) and sSFR. Based on the study by \citet[][see their Fig~7]{schiminovich07}, we divide the parameter space into two-zones corresponding to red and blue galaxies. The galaxy grouping into the two colors following this definition is consistent with the color definition from \citet{tumlinson11b}, which defines galaxies with sSFR higher than $\rm 10^{-11}~yr^{-1}$ as blue and less than this value as red. }
 \label{fig-sample_color} 
\end{figure}

\section{RESULTS\label{sec:results}}

We detected \Lya absorption in the CGM of 36 out of 39 galaxies where measurements could be made. In the remaining 6 sightlines, the data were contaminated by Milky Way or other intervening absorption systems. For the three non-detections, we report a limiting equivalent width that corresponds to 3 times the noise in the spectra in the vicinity of the expected transition. Most of the \Lya features are saturated (zero residual intensity in the line core) with an equivalent width of more than 0.3~$\rm \AA$.

\subsection{Tracing the Circumgalactic Medium}

We begin by assessing the connection of the detected \Lya absorbers to the CGM. Are we seeing gas that is physically connected to the galaxy (a true CGM) or simply material in the inter-galactic medium located close to the galaxy in projection?

\begin{figure*}
\includegraphics[trim = 00mm 0mm 0mm 12mm, clip,scale=0.405, angle=-0]{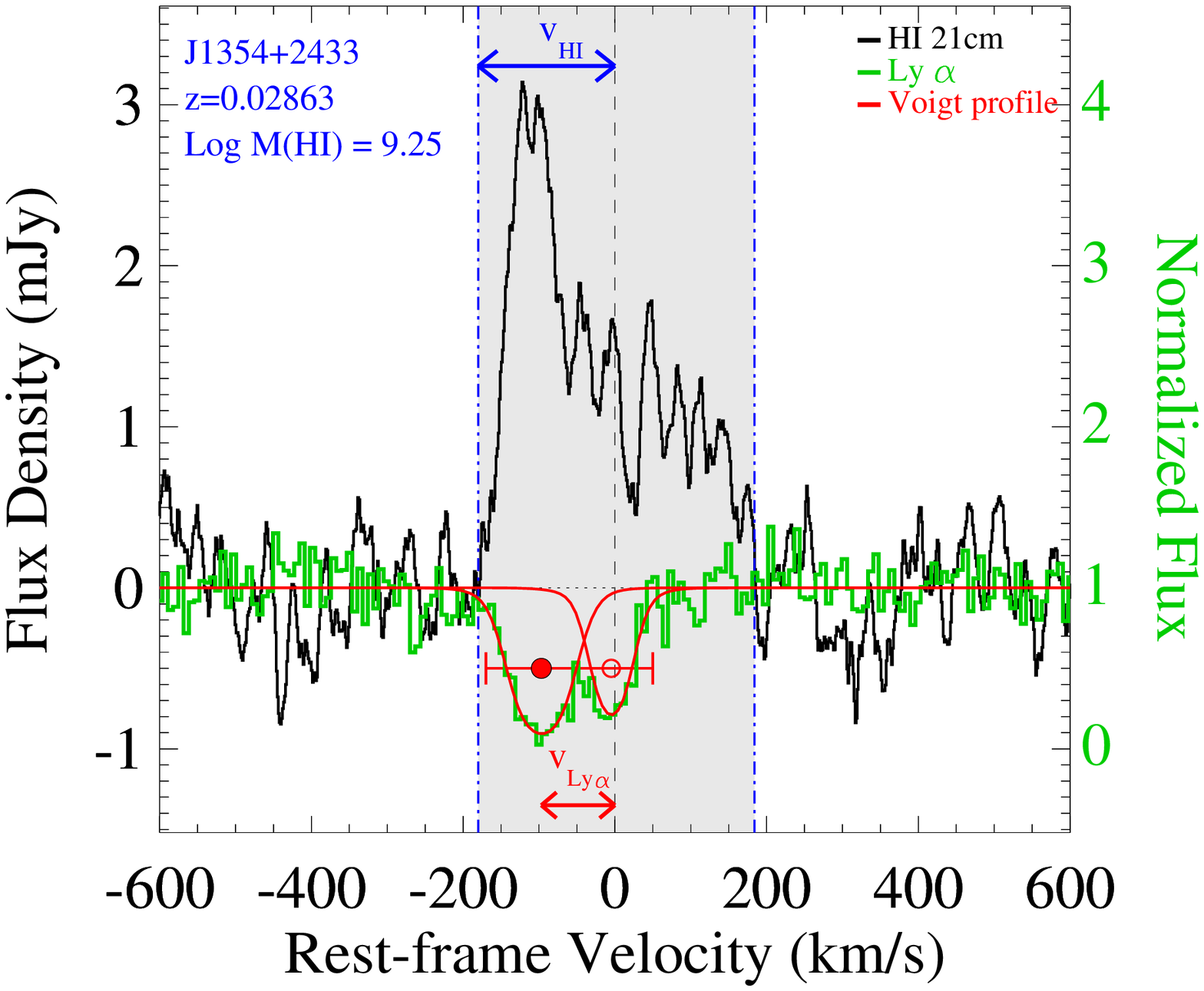}  
\includegraphics[trim = 00mm 0mm 25mm 5mm, clip,scale=0.65, angle=-0]{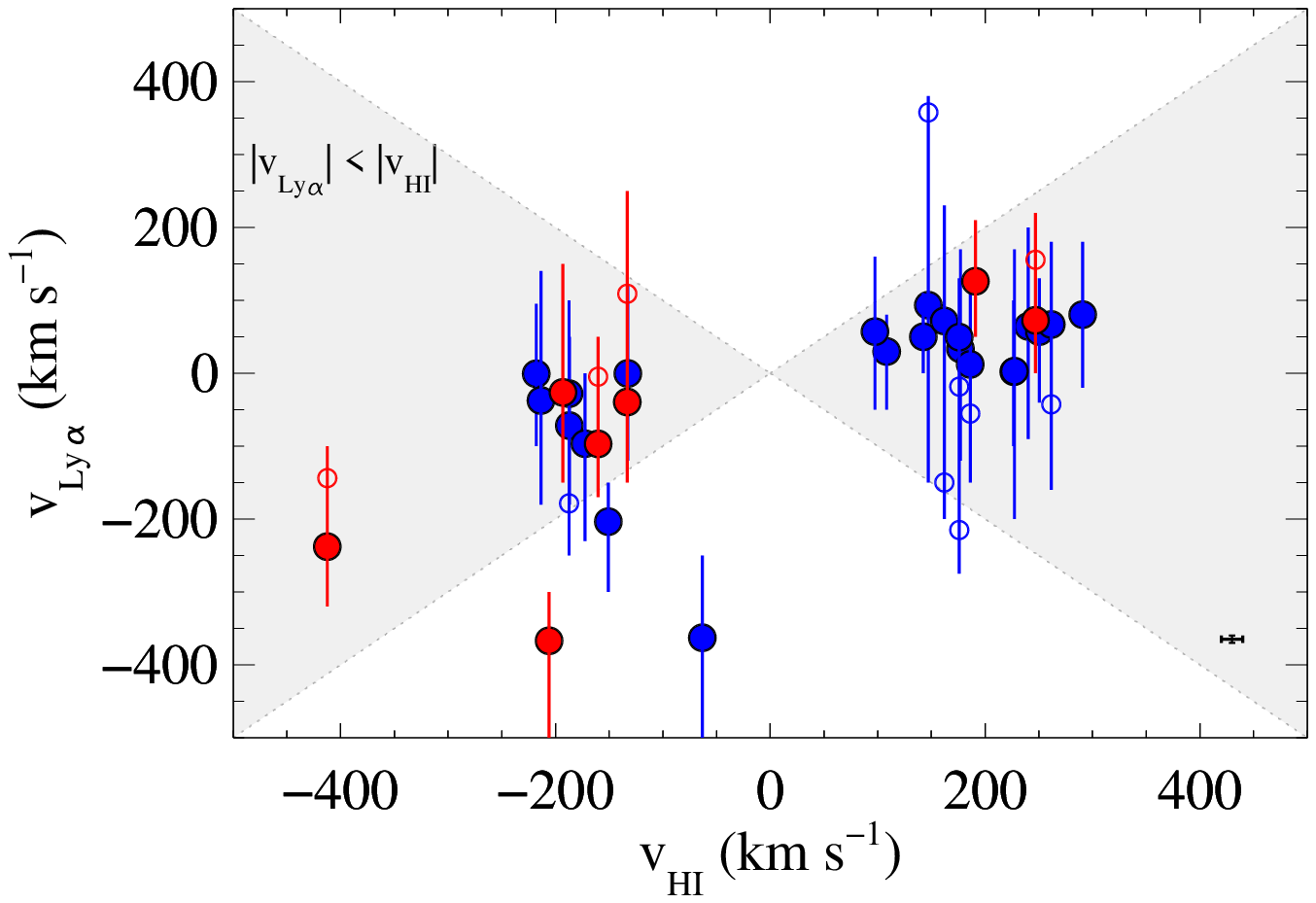}   
\caption{Left: 21~cm \textsc{Hi} spectra of one of our target galaxies ($\rm M_{HI}=1.8 \times 10^9~M_{\odot}$, $\rho=$155~kpc) from the GASS sample. This illustrates a typical \textsc{Hi} profile for our sample. The abscissa and ordinate show the velocity in the rest-frame of the galaxy (based on SDSS emission-line estimation) and the flux in the 21~cm \textsc{Hi} hyperfine transition respectively. The part of the spectrum showing the emission (observed, uncorrected for instrumental broadening) is indicated with the grey shaded region (). The spectrum is smoothed to 20~km/s, and hence its resolution. The normalized \Lya profile for the galaxy is shown in green. The velocity centriod of the strongest component of the \Lya profile is shown as the filled red circle and that of the weaker component is shown as red open circle. The entire width of the \Lya profile is shown as the red line passing through the circles representing the centroids. The velocity offset of the  centroid of the primary \Lya component from the systemic velocity (at 0 \kms in the plot) is termed as $ v_{Ly\alpha}$. We compare this with the velocity extent (half) of the \textsc{Hi} profile, $v_{HI}$, again with respect to the systemic velocity of the galaxy, towards same velocity direction as the $ v_{Ly\alpha}$. For example, if $  v_{Ly\alpha}>0$, then we define $ v_{HI}$ as the velocity extent of the 21~cm \textsc{Hi} profile towards the positive velocity wing of the profile. For the case shown above, $ v_{HI}$ is defined on the negative velocity wing of the \textsc{Hi} profile and is marked on the plot as the blue double headed arrow. Choosing the right velocity edge of the 21~cm \textsc{Hi} profile to compare with the offset of the \Lya  centroid from the systemic velocity reduces the effect of asymmetries on the \textsc{Hi} profile in the analysis. For symmetric 21~cm \textsc{Hi} profiles, $ v_{HI}$ is simply the velocity half width of the profile. Right: The offset of the velocity centroid of \Lya absorbers $ v_{Ly\alpha}$ as a function of the velocity (half) widths, $ v_{HI}$, of the 21~cm \textsc{Hi} profiles. The velocities are measured with respect to the systemic velocity of the host galaxies based on the redshift measurements from optical emission lines. The \textsc{Hi} velocity widths are corrected for instrumental broadening. The centroid of the strongest component is shown as a filled circle and those of weaker components are shown as open circles. The full widths of the \Lya features are shown as the vertical bars passing the centroids. The color of each circle indicates the color of the galaxy. The shaded grey region indicates the parameter space where the \Lya centroids lie within the \textsc{Hi}  extent of the galaxies. The uncertainties in the velocity measurements are shown in the right hand corner. }
 \label{fig-velocity}
\end{figure*}

First, we can compare the distribution of the \Lya absorbers with respect to the virial radius ($\rm R_{vir}$) of the dark matter halo. Our sightlines probe gas at impact parameters comparable to the $\rm R_{vir}$ (so, we sample mostly the outer region of the halo). In order to cover the full range in impact parameter more completely and with better statistics, we combine our sample with the COS-Halos sample \citep{tumlinson13}. The COS-Halos sample covers a similar range in stellar masses and virial radii as our COS-GASS sample. For consistency, we recalculated the halo masses and virial radii for the COS-Halos sample from their stellar masses using the prescription by \citet{behroozi10}.  Similar to our limiting values, we assign the non-detections in the COS-Halos sample a value equivalent to 3 times the root mean squared (rms) noise in the spectra. The only remaining difference that we do not correct for is the difference in redshift between the two samples of about $\sim$ 0.1. The combined sample together covers a large range of impact parameters from 10-250~kpc. This corresponds to a range in normalized impact parameter ($\rho$/R$_{\rm vir}$) of 0.2-1.4. Thus, the combined sample fully samples the CGM of the galaxies as a function of normalized impact parameter.

We re-investigate the distribution of neutral hydrogen as traced by \Lya equivalent width as a function of impact parameter and the normalized impact parameter for our combined sample. Similar studies have been done in the past with different samples \citep[][and references therein]{lanzetta95,chen98, tripp98, chen01b, bowen02, prochaska11, stoc13, tumlinson13, liang14}, and the \Lya equivalent width was found to decrease with increasing impact parameter.
Figure~\ref{fig-rho_W} shows  the equivalent width of the \Lya absorption line as a function of impact parameter and normalized impact parameter ($\rm \rho/R_{vir}$) for this combined sample. The \Lya equivalent widths are plotted in logarithmic units to show the range of values detected. The COS-GASS sample is plotted as filled circles and the COS-Halos sample is plotted as diamonds. Blue and cyan represents galaxies that have specific star formation rates (sSFR) of higher than $\rm 10^{-11}~yr^{-1}$. These galaxies are also referred to as ``blue" galaxies in the rest of the paper. In red and yellow, are galaxies from the combined sample that have sSFRs lower than that value. These galaxies are referred to as ``red" galaxies hereafter. The color definition is adopted from \citet{tumlinson11b}. However, the color assignments for the COS-GASS sample is also consistent with the more rigorous definition by \citet[][see their Fig 7]{schiminovich07}. The COS-GASS sample on a plot similar to Fig~7 of \citet{schiminovich07} is shown in Figure~\ref{fig-sample_color}. The parameter space is divided half-way into red and blue galaxies.  
However, it is worth noting that galaxies closest to the dividing line are quite diverse and might correspond to a transitional phase (also known as ``green valley"). 
For example, some of these galaxies show the Balmer series transition in emission as well as absorption. They also have a wide range of \Lya strength in their CGM. Unfortunately, our sample does not have a statistically significant number of such transitional phase galaxies so we can not probe the complexities of their CGM in detail.  

The upper panel of Figure~\ref{fig-rho_W} shows that the equivalent widths of the \Lya absorbers start to drop off beyond an impact parameter of about 150 kpc. This same behavior is seen in the plot {\it vs.} normalized impact parameter (lower left panel). An alternative way to visualize the radial distribution of the \Lya absorbers is shown in the lower right panel, where the normalized impact parameter is plotted in linear units. A linear fit to the data in this plot yielded the following best-fit parameters
\begin{equation}{\label{eq-LogW_lya}}
\rm Log W_{Ly\alpha}= (-0.432\pm 0.092) \frac{\rho}{R_{vir}} ~-~(0.026\pm 0.039)
\end{equation}

This implies that the strength of \Lya data can be described as an exponential function of normalized impact parameter with an e-folding length of $\rm R_{vir}$ as follows:
\begin{equation}{\label{eq-W_lya}}
\rm W_{Ly\alpha}= 0.941~e^{~-\rho/R_{vir}} 
\end{equation}

This same fit is shown in the lower left panel. A similar fit using the (un-normalized) impact parameters is shown in the top panel and yields a scale-length of 136 kpc.

\begin{figure*}
\includegraphics[trim = 0mm 0mm 0mm 0mm, clip,scale=0.5,angle=-0]{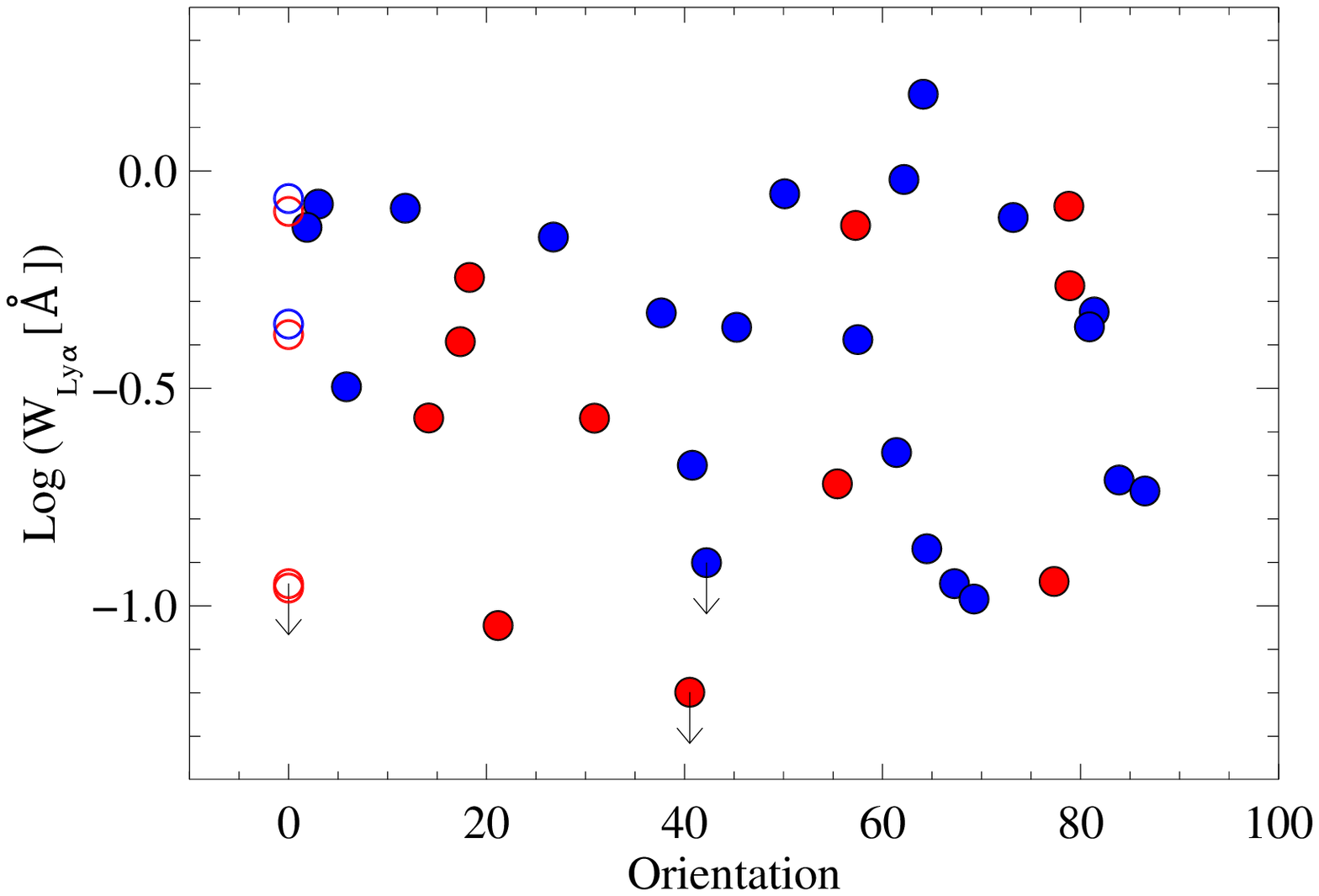}  
\includegraphics[trim = 0mm 0mm 0mm 0mm, clip,scale=0.5,angle=-0]{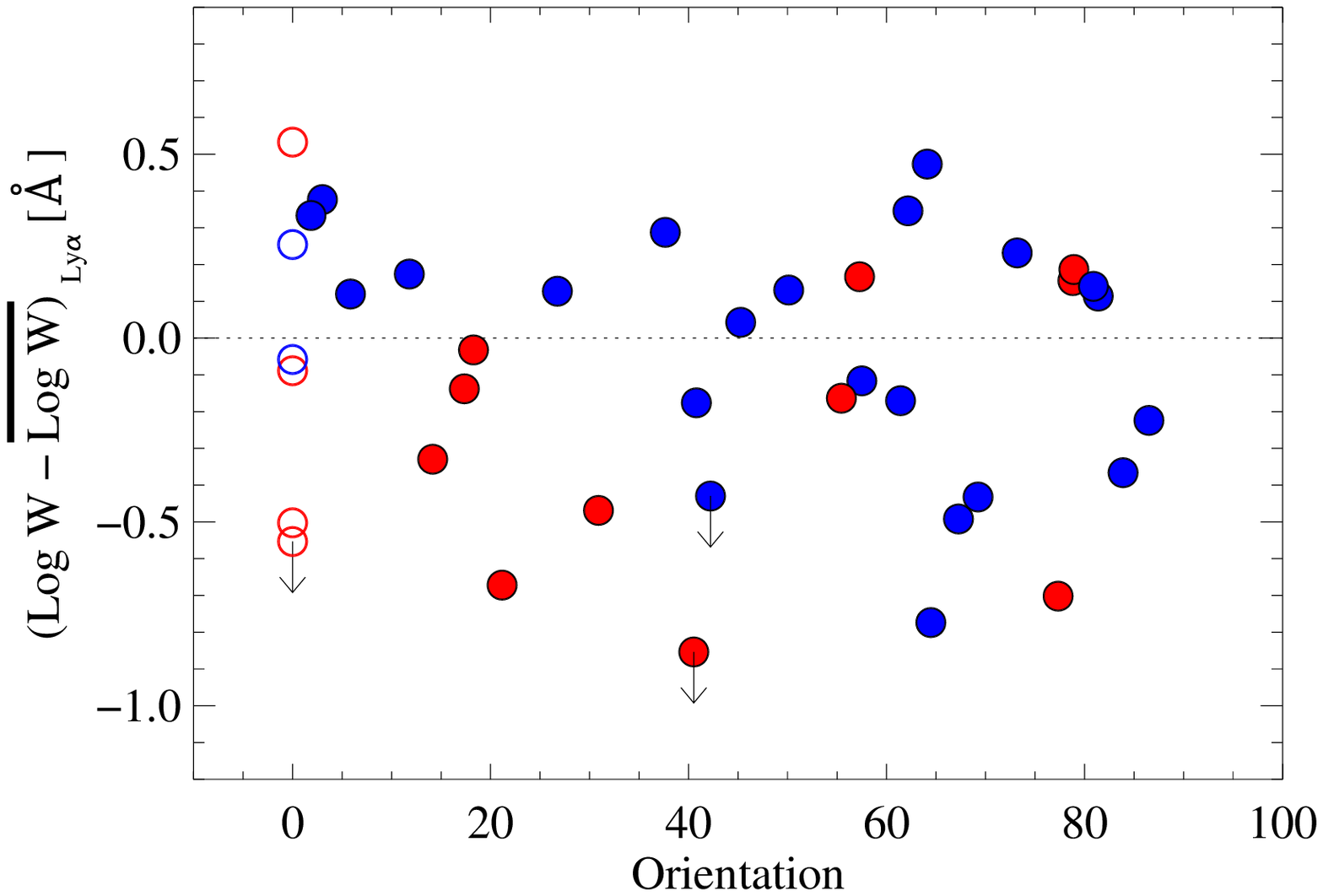}  \\
\caption{Left: \Lya equivalent width as a function of orientation of the QSO sightline with respect to the optical major axis of the galaxy. The open circles represent galaxies that are face-on and their orientation parameter is set to zero. The color of the circles represents the color of the galaxy that is summarized in Table~1. Right: The excess in \Lya equivalent width ([Log W - $\rm \overline{Log~W}]_{Ly\alpha}$), which is independent of the impact parameter of the sightlines, as a function of orientation of the QSO sightlines with respect to the target galaxy. The dotted line corresponds to the expected equivalent width based on fit to Figure~2. We find no correlations between orientation and \Lya strengths. The face-on galaxies are excluded while measuring the strength of the correlation.}
\label{fig-Orientation} 
\end{figure*}

We can also assess the velocity offsets between the \Lya absorbers and the systemic velocity of the central galaxy. Is this gas likely to be bound to the galaxy? We have therefore measured the displacement of the centroid of the \Lya absorption line from the galaxy's systemic velocity. The systemic velocity is defined as the velocity corresponding to the SDSS-based redshift measurement i.e. $\rm V_{sys}= \it ~c~z$, where $c$ is the speed of light in vacuum. The uncertainty in systemic velocity is of the order $\sim$30~\kms.

Among the sightlines where \Lya was detected, 34/36 (94\%) of the absorption features showed the centroid of the strongest component of the \Lya feature to be within 250~\kms of the  galaxy's systemic redshift. This suggest that the absorbers are most likely bound within the dark matter halo of the galaxy.  To investigate this further, we estimate the escape velocity, for a parcel of gas located at the (1) distance of closest impact of the sightline  ($\rm v_{esc}^{\rho}$) and  (2) virial radius ($\rm v_{esc}^{Rvir}$). The values are presented in Table~2. These calculations were made assuming a spherically symmetric NFW profile \citep{nfw} for the dark matter halo with a concentration parameter of 15.
Only 2/36 of \Lya absorbers have the velocity centroid of the strongest component larger than either of the escape velocities. This strongly suggests that the bulk of the \Lya absorbers are tracing the bound CGM. 
Similarly, if we consider all the components of the \Lya profiles, then the number of possibly escaping components increases to 5/48.
In summary, 34/36 (94\%) of the strongest \Lya components or 43/48 (90\%) of all \Lya components have velocities smaller than escape velocity. Therefore, statistically the \Lya absorbers in our sample are likely to be probing the bound CGM.

We also find that the \Lya absorption-lines show velocities consistent with that of the velocity range observed in the 21cm line profiles of the \HI gas within the galaxies. 
 An example of the 21cm \HI profiles of one of our galaxies (J1354+2433) is presented in the left panel of Figure~\ref{fig-velocity}. The Arecibo spectrum of the galaxy shows \HI spread over a $\approx \rm \pm 200$~\kms (marked as the shaded region) in the rest frame of the galaxy. The rest-frame velocity was determined by subtracting the systemic velocity (v$_{sys}=c~z$) from the observed \HI velocities.  
The velocity centroid of the \Lya absorption feature ($\rm v_{Ly\alpha}$) with respect to the systemic velocity of the host galaxy is marked as the red filled (strongest component) and open (weaker) components. The maximum velocity extent of the 21cm \HI line profile in the same velocity end as the \Lya centroid is defined here as $\rm v_{HI}$. The need to measure $\rm v_{HI}$ on either the redshifted or blueshifted side of the spectra based on the location of the \Lya centroid arises due to the fact that 21~cm \HI profiles may be asymmetric with respect to the systemic velocity. For instance, if the \Lya centroid is blueshifted in the rest-frame of the galaxy, then we measure $\rm v_{HI}$ as the maximum velocity of the 21cm \HI profile at zero flux blueward of the rest-frame. Hence, the estimates of the velocity offsets of the ISM and the CGM at the rest-frame are not impacted by the asymmetries in the 21cm \HI profile.
 
The right panel of Figure~\ref{fig-velocity}  shows the rest-frame velocity distribution of the \Lya absorbers as a function of velocity distribution of 21cm \HI profiles that traces the ISM of the host galaxies. In this figure, the centroids of the \Lya absorption features are marked with circles and the full range of velocity for the absorption features are shown as vertical bars. The centroid of the strongest component is shown as filled circle and the weaker components, if present, are shown as open circles. The color of the symbol represents the color of the galaxy as summarized in Table~1. The velocities are normalized to the rest-frame using optical redshifts from the SDSS.

The grey shaded area represents the region of the space where $\rm |\delta V_{Ly\alpha}| \le |\delta V_{HI}|$. 90\% (26/29) of the centroids of the strongest \Lya absorbers lie within the velocity range of the 21~cm \HI emission-line associated with the target galaxy. Two notable exceptions are the blue galaxy J1541$+$2817 and the red galaxy J1515$+$0701. The \HI spectral range in J1541$+$2817 extends from $\rm -73~to~107$~\kms and the \Lya absorber was detected at -399~\kms. Similarly, the \Lya velocity centroid of J1515$+$0701 is -362~\kms whereas the \HI associated with the galaxy spans a range of $\rm -207~to~303$~\kms with respect to the systemic velocity of the galaxy. The origin of the offsets in these two systems is not clear (they do not seem to be in unusual environments relative to the other systems). For most galaxies, that is 20/29 ($>$66\%), the entire \Lya profile lies within the 21cm \HI profile. Therefore,we conclude that the vast majority of the absorbing systems are very likely to be gravitationally bound material inside the dark matter halo.
 
Next we have looked for any dependence in the properties of absorbers on the orientation of the sightline with respect to the major axis of the galaxy. This would potentially allow us to distinguish between \Lya arising in an extended disk, in bipolar outflows, or in material simply tracing the dark matter halo.  Figure~\ref{fig-Orientation} presents the strength of \Lya equivalent width as a function of orientation of the sightline with respect to the galaxy \citep[e.g][]{kacprzak10,bordoloi11}.  No correlation has been observed.
Since we showed previously that \Lya equivalent widths correlate with the normalized impact parameters, we corrected for any impact parameter dependence by dividing the observed \Lya equivalent width by the average \Lya equivalent width at that impact parameter (described by the best-fit line of Equation~\ref{eq-LogW_lya}).  We refer to this term as the 'impact parameter corrected excess in equivalent width', or simply as the 'excess equivalent width'.  Here again, we see that the excess in the \Lya equivalent width does not correlate with orientation of the sightline with respect to the  major axis of the target galaxy for the full sample. Part of the reason for the non-existence of any correlation could be because we are probing mostly the outer CGM.

{\it Therefore, we conclude that the \Lya absorbers we are studying trace a roughly spherical distribution of gravitationally-bound gas with a characteristic size comparable to the virial radius. This is what we would expect for the CGM.}

\begin{figure*}
\includegraphics[trim = 20mm 127mm 20mm 0mm, clip,scale=0.50,angle=-0]{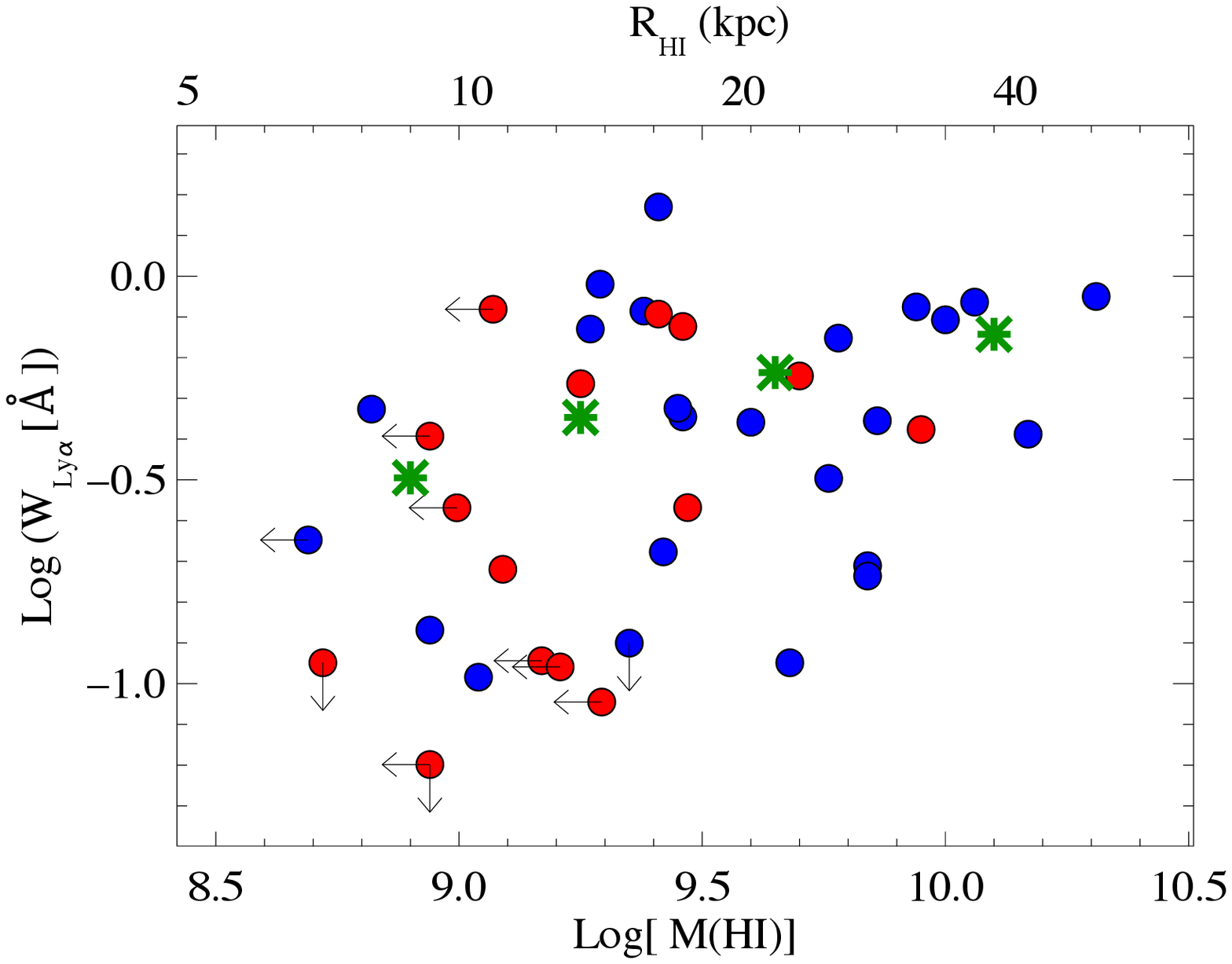} 
\includegraphics[trim = 00mm 000mm 00mm 0mm, clip,scale=0.50,angle=-0]{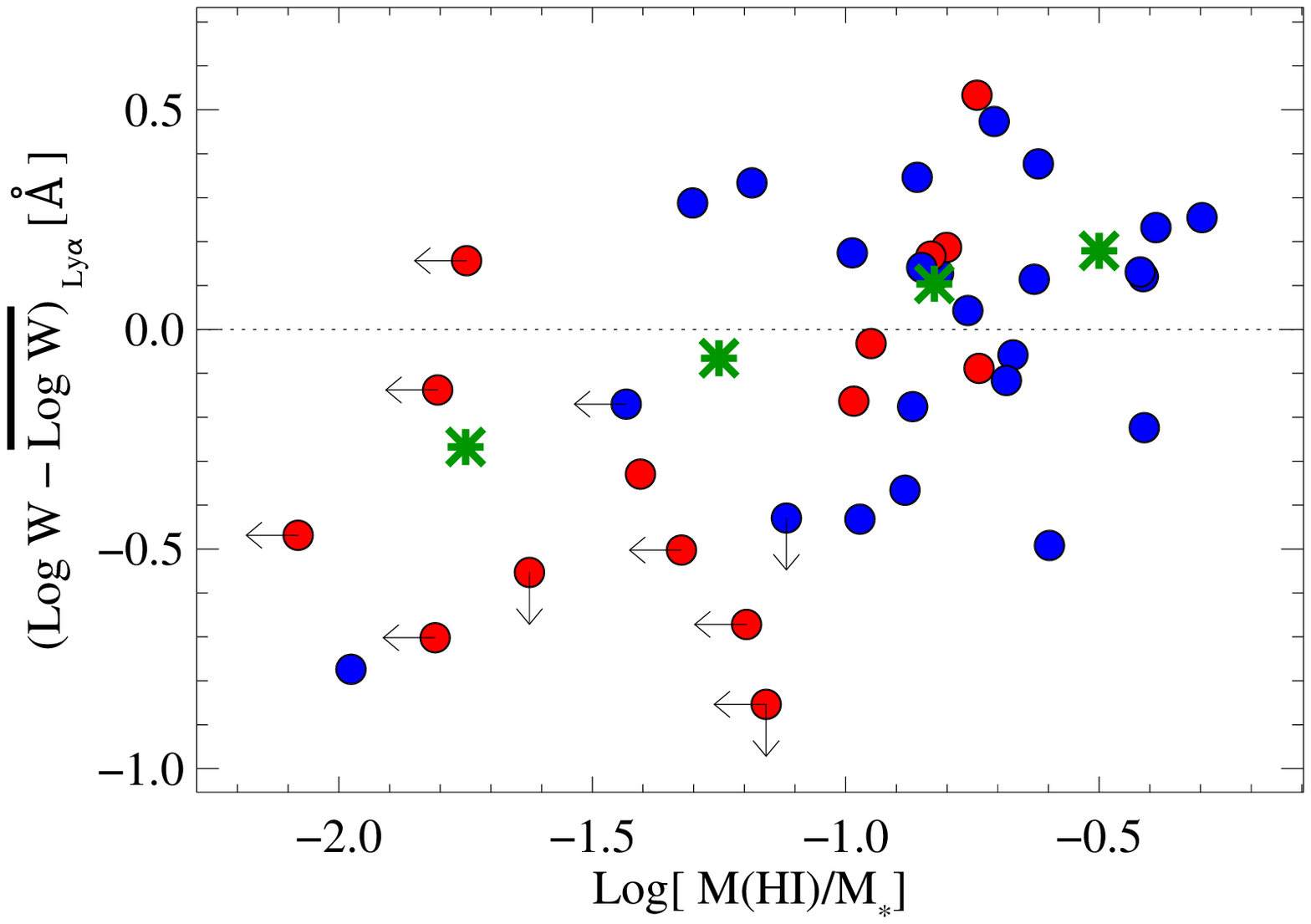} 
\caption{Equivalent width of \Lya as a function of 21~cm \textsc{Hi} mass and the excess in \Lya equivalent width ([Log W - $\rm \overline{Log~W}]_{Ly\alpha}$), which is independent of impact parameter of the sightline, as a function of the \textsc{Hi} mass fraction $\rm M_{HI}/M_{\star}$ of the galaxies. The size of the \textsc{Hi} disk is indirectly estimated from the tight \textsc{Hi} size-mass relation \citep{swaters02}. We find statistically significant correlations with a statistical significance of 98.5\% for the left panel and 99.8\% for the right. The top x-axis of the left panel converts the \textsc{Hi} mass to a radius of the \textsc{Hi} disk (see text for details). The average equivalent widths derived from the stacked spectra (presented in the next figure) are plotted as green asterisk in the left panel. The averages derived from the data points in the right panel are shown as green asterisk.} 
\label{fig-HI} 
\end{figure*}

\subsection{Correlation between properties of the CGM and the ISM \label{sec:ISM_CGM} }

We now explore the connection between the gas in the CGM as traced by the \Lya absorbers and the cool neutral ISM as traced by the HI 21cm line.
Figure~\ref{fig-HI} shows the results for the COS-GASS sample. The abscissa indicates their 21~cm \HI mass on the bottom and their estimated \HI radii on the top. The radius of the disk, $\rm R_{HI}$ \footnote{$\rm R_{HI}$ is defined as the annuli of \HI mass density of    $\rm 1~M_{\odot}~pc^{-2}$ or column density $\rm 1.3\times 10^{20}~cm^{-2}$.}, was indirectly estimated using the extremely tight relationship between \HI mass and size as observed by \citet{swaters02}. 
We find a positive correlation between the equivalent width of \Lya absorbers in the CGM and $\rm M(HI)$.  A Kendall $\tau$ \citep{brown74} test implies that the null hypothesis of no correlation can be rejected at the 98.5\% confidence level$\rm ^{4}$. This also implies a positive correlation between the equivalent width of \Lya and the size of the \HI disk. We find even stronger correlations between the suitably normalized quantities\footnote{We are using $\rm M_{\star}$ as a proxy for halo mass. The normalization by the two kinds of masses yield very similar results as $\rm M_{halo}$ is derived from  $\rm M_{\star}$. }. This is shown in the right panel where the excess \Lya equivalent width is plotted as a function of \HI mass fraction ($\rm M(HI)/M_*$) of the host galaxy. The resulting correlation is inconsistent with the null hypothesis at the 99.8\% confidence level\footnote{The test was performed on our censored data using the astronomy survival analysis code ASURV \citep{ASURV}.  ASURV is capable of handling single and doubly censored data. Accuracy of these probabilities can be affected by larger number of censored values and other conditions. Since less than a quarter of our sample has censored values, we do not expect substantial inaccuracies. However, caution is appropriate as is for results from Kendall's test  on any sample \citep{wang00}.}. At the same time, we do not find any correlation between the \Lya equivalent width and the stellar mass of the galaxies.

\begin{figure*}
\includegraphics[trim = 0mm 0mm 0mm 0mm, clip,scale=0.6,angle=-0]{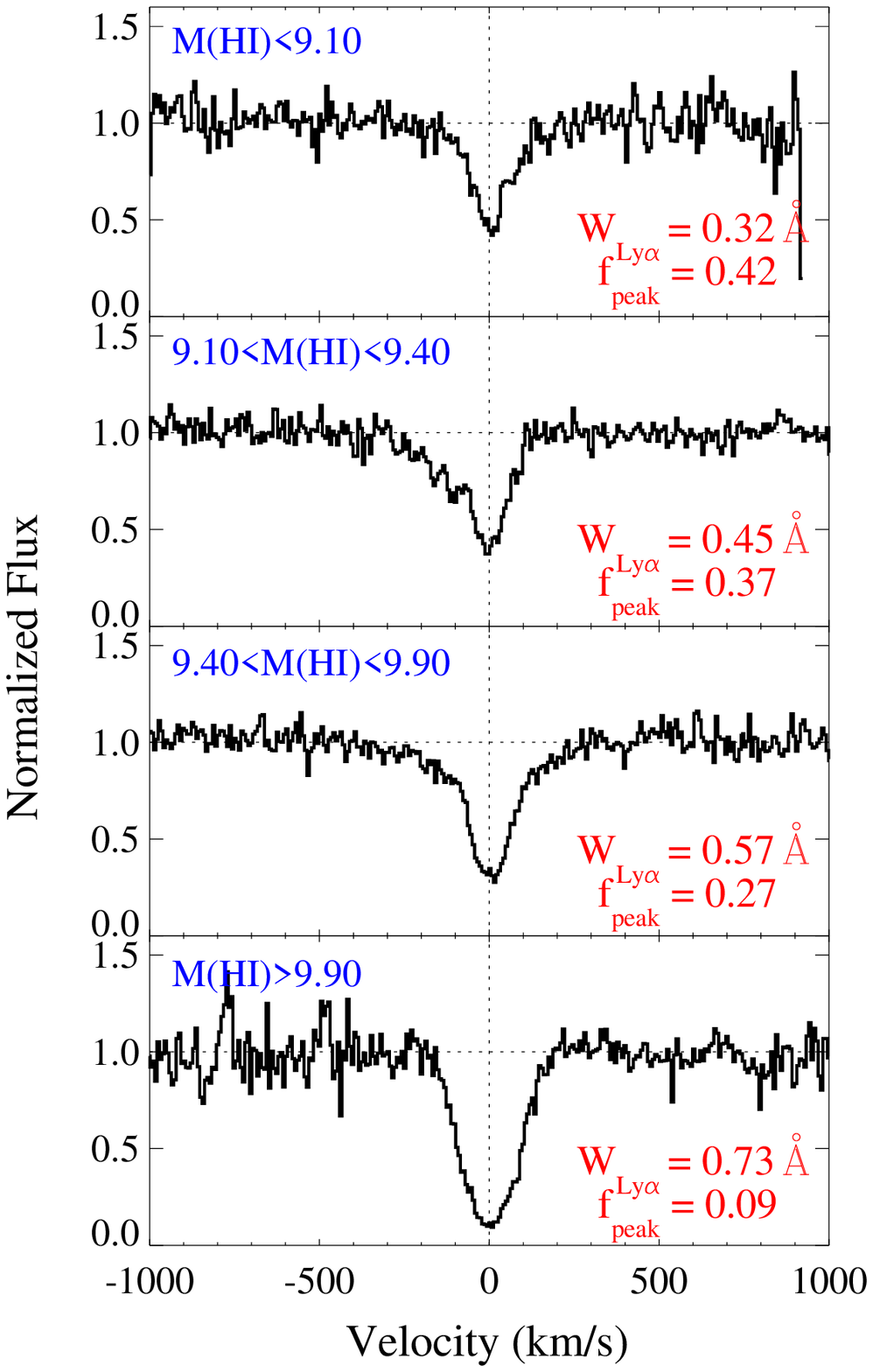} 
\includegraphics[trim = 0mm 0mm 0mm 0mm, clip,scale=0.6,angle=-0]{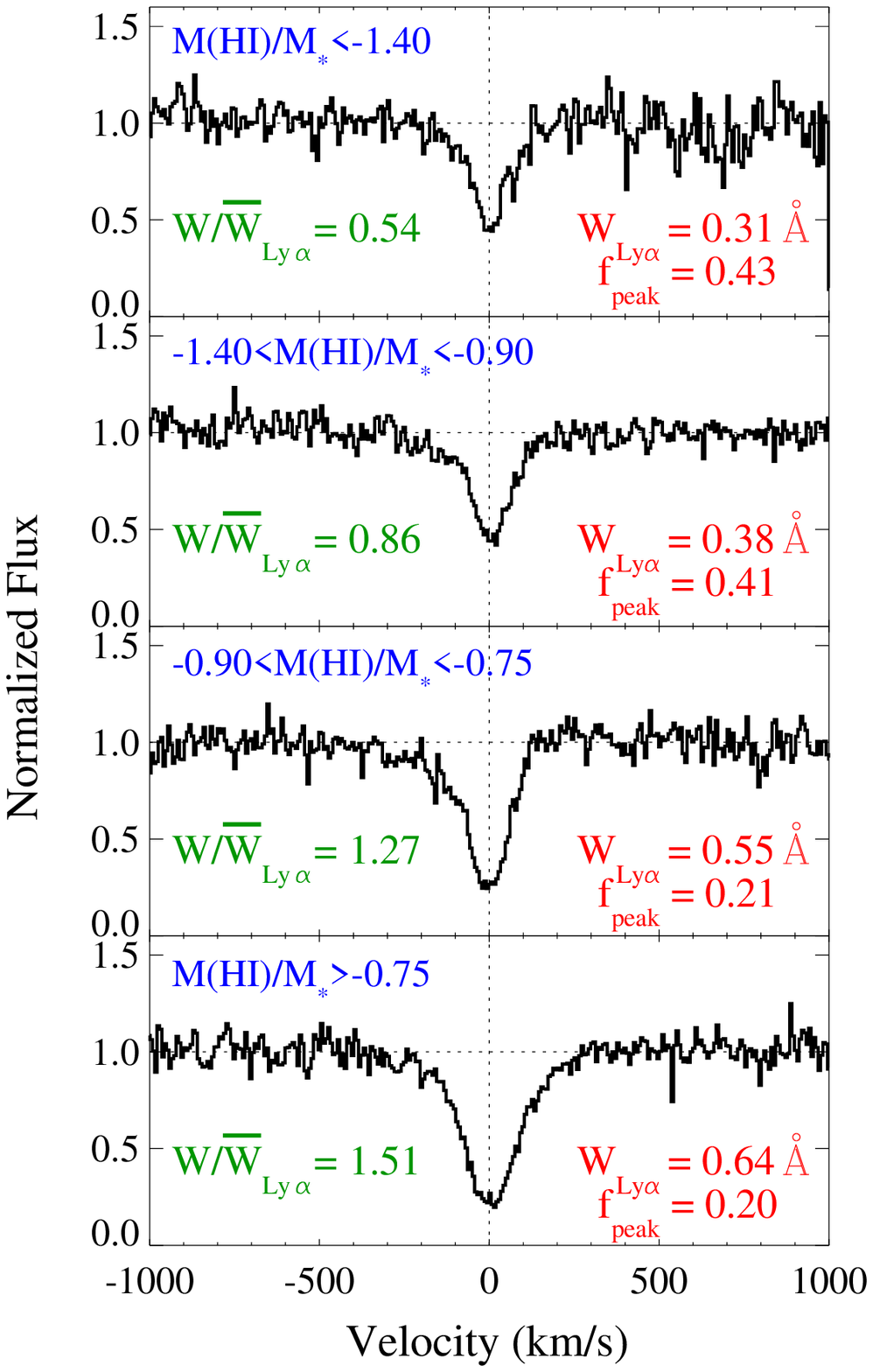} 
\caption{Composite \Lya absorption-lines stacked as a function of \textsc{Hi} mass and gas fraction ($\rm M(HI)/M_{\star}$). Individual \Lya absorbers were centered at their centroid of the feature i.e. $\rm v_{centroid}=0$~\kms,  and then added to produce the stacks. Each of the stacked spectra are composed of approximately 8-10 individual spectra in that labeled \textsc{Hi} mass range. The equivalent width and the peak depth (f$\rm_{peak}$) of the absorption features are labeled in the right corner of each plot in red. The average values of the excess in \Lya equivalent width  for the sightlines used in creating the stacks are printed in green. These set of stacks are not corrected for impact parameter variations and consequently show a weaker trend just like the left panel of Figure~5. Similar plots with impact parameter bins are presented in the next figure. }
\label{fig-HIstacks} 
\end{figure*}

To visualize our data and to help in its interpretation, we also performed a stacking experiment to study the average properties of the CGM as a function of 21~cm \HI mass within their host galaxies.  We divided our sample into four sub-samples based on their \HI masses. Spectra of each sub-sample were stacked within $\pm$1500~\kms of the rest-frame \Lya transition associated with the target galaxy. Each spectrum was centered at the centroid of the \Lya absorption feature before the stacking was performed. For the three non-detections, we used the systemic velocity of the target galaxies as zero velocity.  The results are presented in Figure~\ref{fig-HIstacks}. Each of these stacks comprise approximately 8-10 spectra.

We detected \Lya absorption-lines in each stack. However, their strength shows considerable variations. The \Lya strengths increased monotonically with increasing \HI mass within the galaxy. Both the peak depth of the absorption features as well as equivalent widths of the absorption features increase with increasing \HI mass. Since most of the individual \Lya absorption-lines are saturated, the behavior of the stacks is indicative of the increase in the covering fraction of \Lya in the CGM of galaxies as a function of increasing \HI mass. A similar trend is also seen as a function of \HI fraction ($\rm M(HI)/M_{\star}$). Another way to illustrate the same conclusion is to note the fractional increase in saturated \Lya lines as a function of \HI mass. For example, for galaxies with  M(HI) $< \rm 10^{9.1}~M_{\odot}$ the fraction of \Lya absorbers with W$_{Ly\alpha} > 0.3\rm \AA$ is about 30\%. It increases to $\sim$55\% and 67\% for galaxies with M(HI) between $\rm 10^{9.1-9.45}~M_{\odot}$ and M(HI) between $\rm 10^{9.45-9.9}~M_{\odot}$ respectively. It reaches 100\% in the sub-sample with M(HI) $> \rm 10^{9.9}~M_{\odot}$.

\begin{figure}
\hspace{-0.75cm}
\includegraphics[trim = 0mm 0mm 0mm 0mm, clip,scale=0.62,angle=-0]{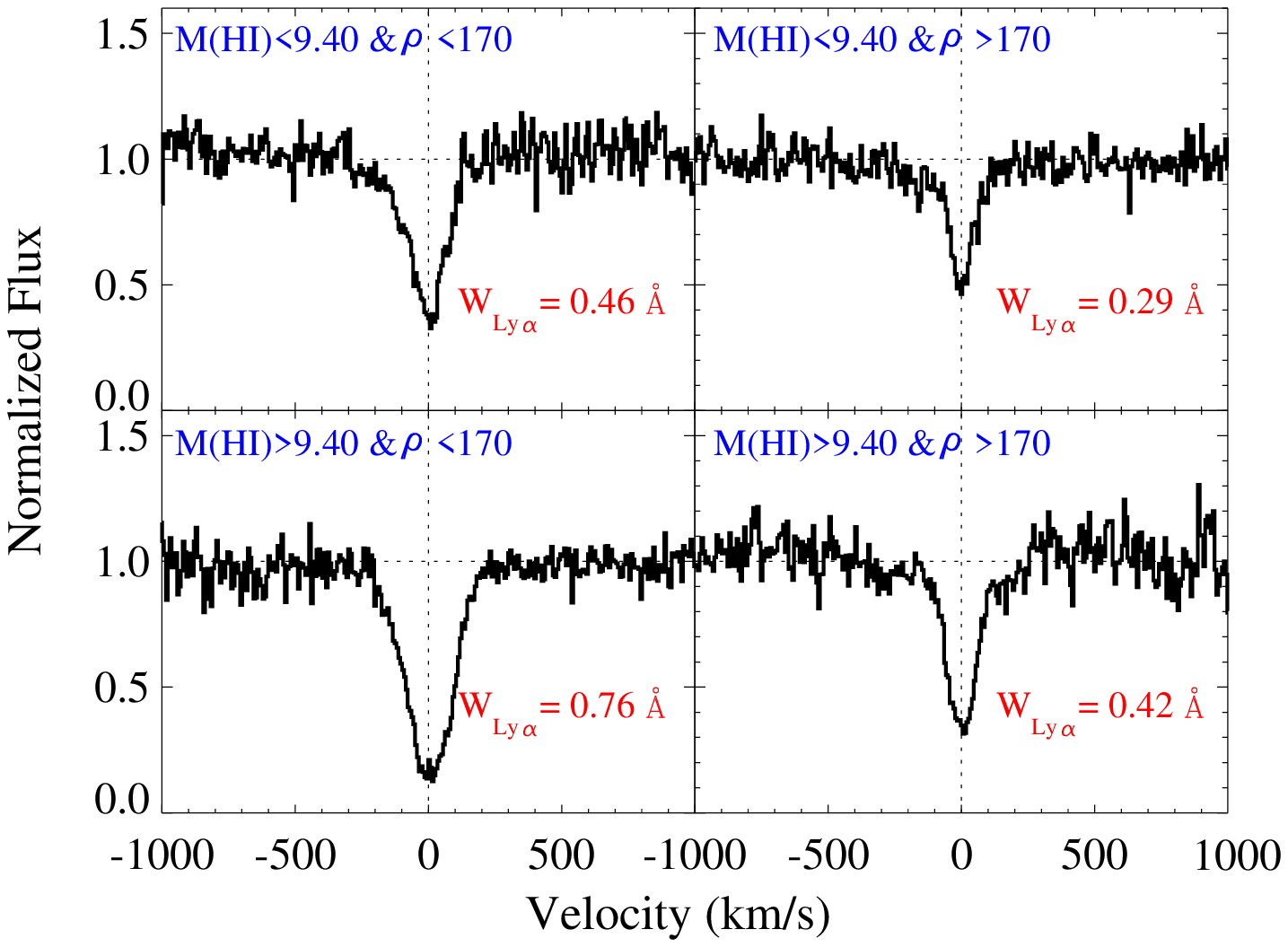} 

\hspace{-0.75cm}
\includegraphics[trim = 0mm 0mm 0mm 0mm, clip,scale=0.62,angle=-0]{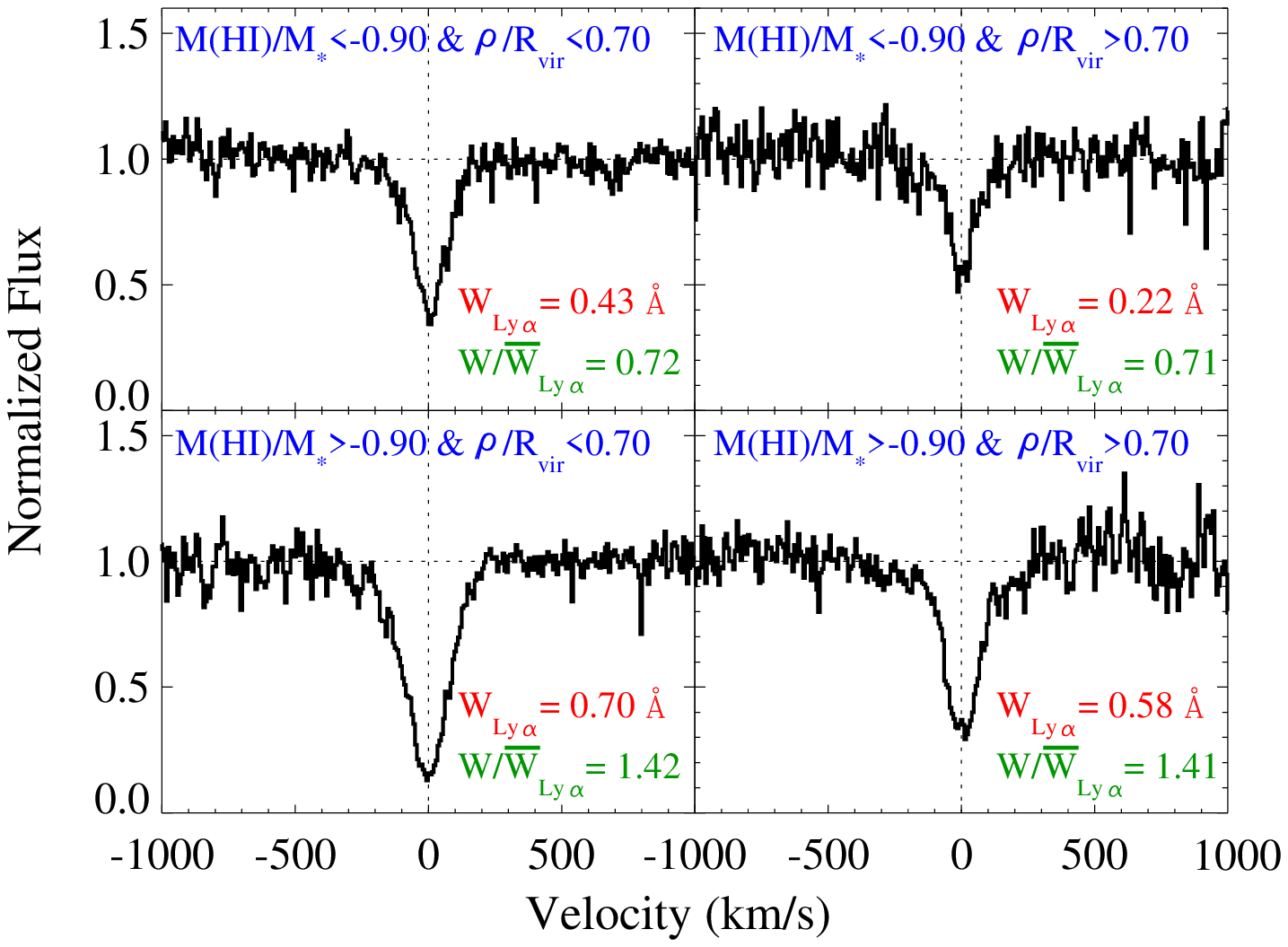} 
\caption{Top: Stacks of \Lya absorption similar to those presented in Figure~\ref{fig-HIstacks} but with bins that divide the sample into subsets of impact parameter - one less than and the other greater than 170~kpc. The values for the bins were chosen such that the sample is divided more or less equally into 4 bins. Systematic variations the strength of the \Lya absorption as a function of \textsc{Hi} mass within each bin can be seen. The equivalent width of the features are labeled in the right corner of each plot in red.  Bottom: Stacks similar to those on the top, but binned by \textsc{Hi} mass fraction and normalized impact parameter ($\rho/\rm R_{vir}$). The sample was divided along the median value of normalized impact parameter of 0.70. The average values of the excess in \Lya equivalent width for the sightlines that go into the stacks are printed in green. As expected, these are independent of the normalized impact parameter bins.}
\label{fig-HIstacks_sub} 
\end{figure}

We also investigated the combined influence of impact parameter and \HI mass on the equivalent width of the \Lya profile. To do so, we divided the sample into 2 impact parameter bins - one  lower  and the other  higher than 170~kpc. We then sub-divided each sub-sample into galaxies with \HI mass less or greater than 9.45~dex. The left panel of Figure~\ref{fig-HIstacks_sub} shows the stacked spectra for each of the bins. This verifies the influence of \HI mass as well as impact parameter separately on the strength of \Lya profiles. A similar set of stacks were produced for the hybrid parameters  - $\rm M(HI)/M_{\star}$ and $\rm \rho/R_{vir}$ - these further confirm the correlation between $\rm M(HI)/M_{\star}$ and the strength of the \Lya absorption lines.

\begin{figure}
\hspace{-1.5cm}
\includegraphics[trim = 0mm 130mm 0mm 20mm, clip,scale=0.5,angle=-0]{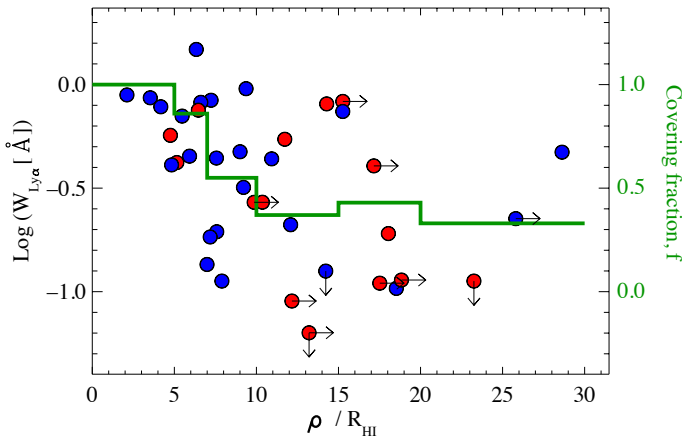}  
\caption{\Lya equivalent width as a function of the impact parameter  normalized by the radius of the \textsc{Hi} disk ($\rm \rho/R_{HI}$). The absorbers are shown as colored filled circles with their colors indicating the color of the galaxy as defined by their sSFR and presented in Table~1. Sightlines within about 7 $R_{HI}$ all produce strong saturated \Lya lines. In green we plot the covering fraction of strong \Lya absorbers with equivalent width larger than $\rm 0.32~\AA $~(-0.5~dex). This corresponds to a column density of  $\rm \approx 6 \times 10^{13}~cm^{-2}$ or larger. The right ordinate is labeled to indicate the scale.}  
 \label{fig-RHI} 
\end{figure}

We also explore the effect of the size of the \HI disk on the properties of the \Lya absorbers.  Figure~\ref{fig-RHI} plots the equivalent widths of the \Lya absorbers as a function of the impact parameter normalized by the HI radius (i.e. $\rho/ \rm R_{HI}$). Sightlines probing the CGM within $\sim$7~$\rm R_{HI}$ all show strong saturated \Lya absorption features ($\rm W_{Ly\alpha}>0.4\AA$). The other striking feature of the plot is the steep drop in the strength of the absorption-lines at larger normalized radii. 

{\it In conclusion, there is a significant correlation between the strength of the \Lya absorption-lines tracing the CGM and the mass and by extension the extent of the distribution of 21~cm \HI component of the galaxy disk.}

\subsection{Relationships of the CGM to the \HI and Star Formation}

While the properties of the CGM traced by \Lya correlate with the \HI properties of the host galaxy, the \HI properties also correlate with the star formation rates of the galaxies  \citep{schiminovich10, catinella10}. In this subsection we examine the correlations between these parameters in an attempt to isolate the primary physical correlations.  This can be addressed most easily by determining if there is a correlation between the \Lya equivalent width and the star-formation rate, and then comparing this to the strength of the correlation between \Lya equivalent width and \HI mass. This will help us probe the physics behind the correlations and isolate the driving mechanisms that connect gas at distances of hundreds of kpc to the central galaxy. 

\begin{figure*}
\includegraphics[trim = 0mm 0mm 0mm 0mm, clip,scale=0.5,angle=-0]{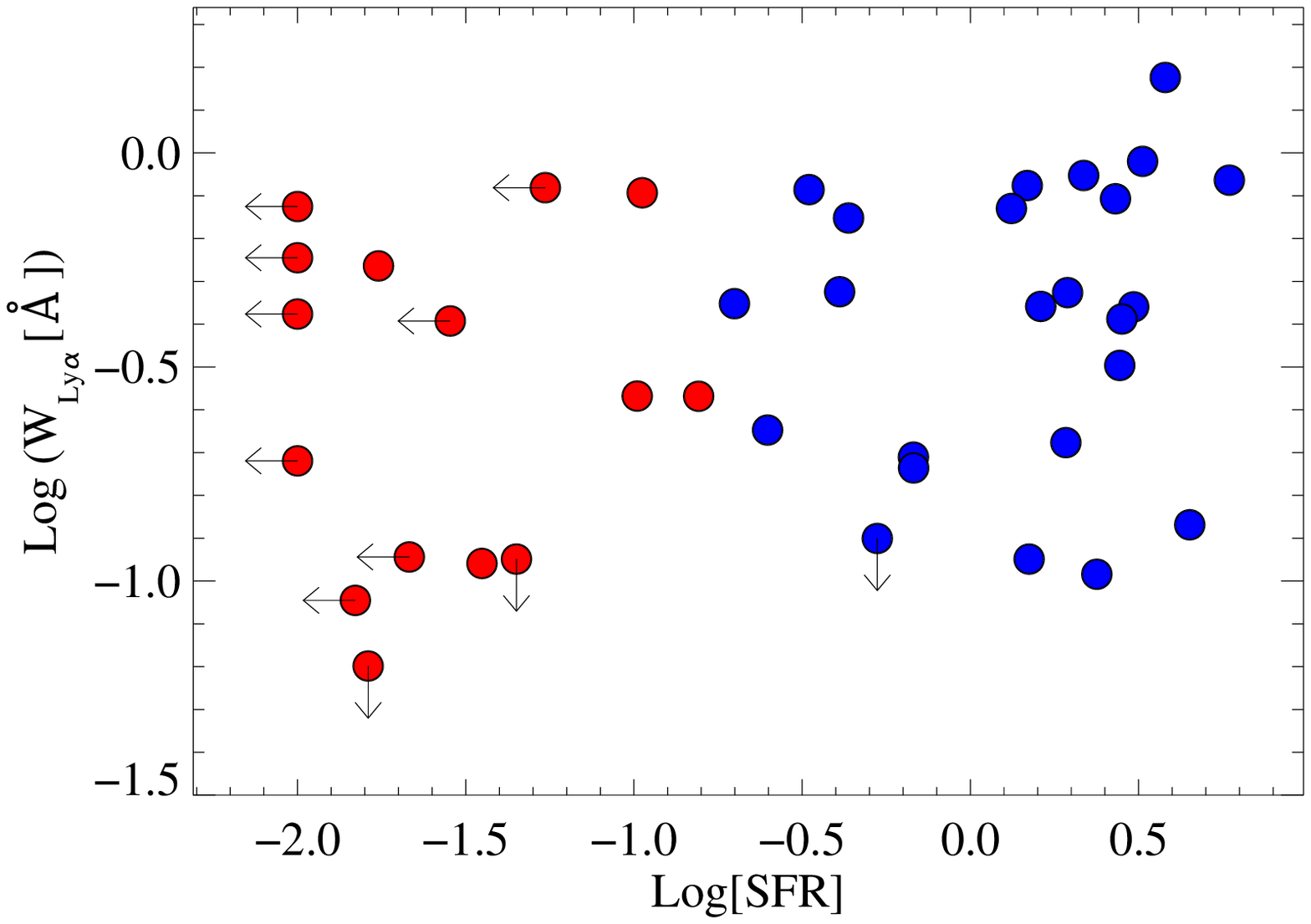}  
\includegraphics[trim = 0mm 0mm 0mm 0mm, clip,scale=0.5,angle=-0]{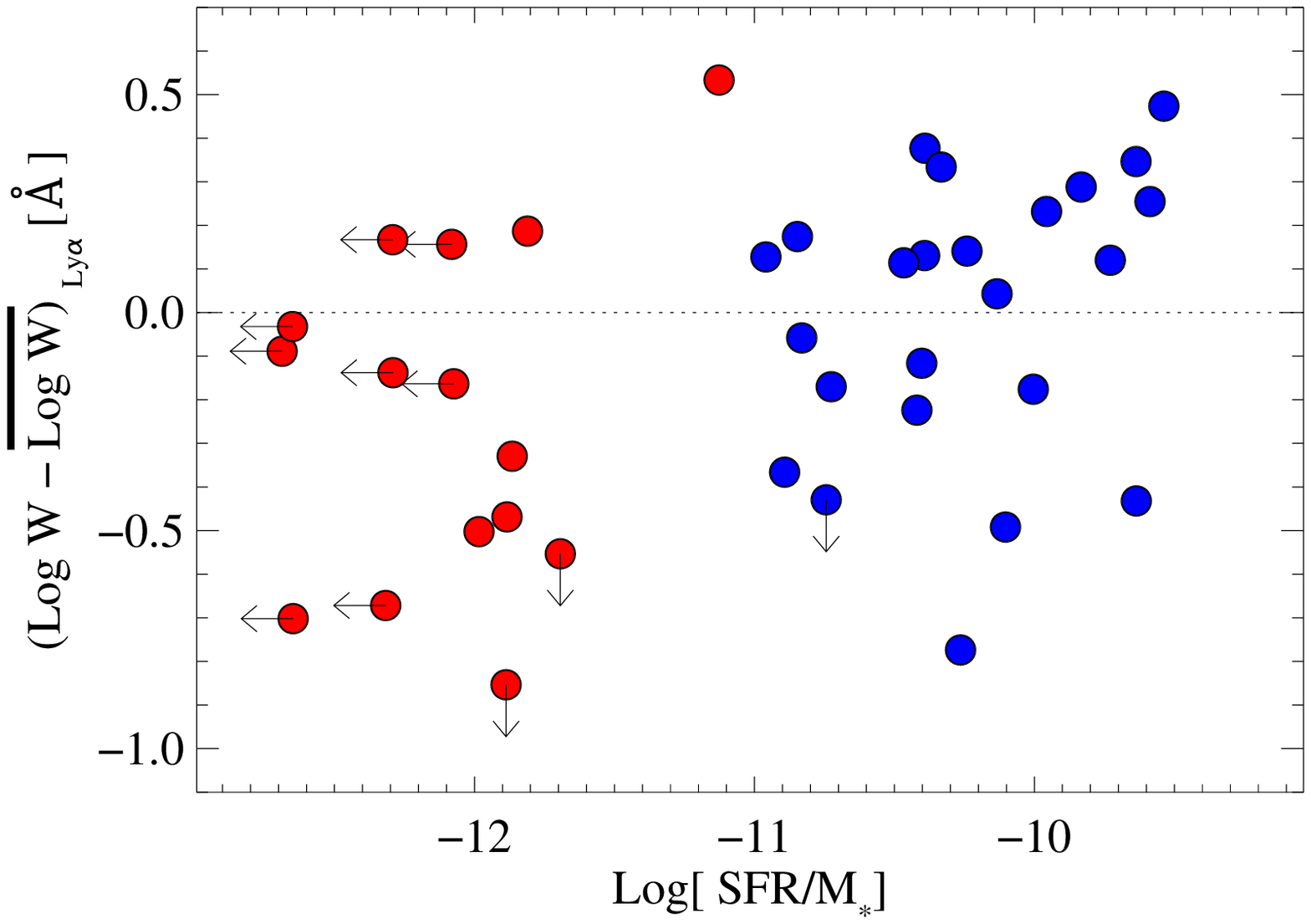}   
\caption{Left:  Variation in the \Lya equivalent width as a function of SFR.  The absorbers are shown as colored filled circles with their colors indicating the color of the galaxy as defined by their sSFR and presented in Table~1. Right: The excess \Lya equivalent width as a function of specific SFR. The correlations have a statistical significance of only 89.9\% (left panel) and 97.5\% (right panel). These are weaker than the corresponding ones in Figure 5.}
 \label{fig-SFR} 
\end{figure*}

We therefore repeat the same exercise for the SFR as was done for the \HI masses in Figure~\ref{fig-HI}. The corresponding plots showing the distribution of \Lya equivalent width as a function of SFR and the excess in \Lya equivalent width as a function of specific SFR (sSFR = SFR/M$_*$) are presented in Figure~\ref{fig-SFR}. The Kendall's rank correlation test reveals that the probability of the null hypothesis of no correlation between \Lya equivalent width and SFR can only be excluded at the 89.9\% confidence level, while the corresponding probability for the correlation between excess \Lya equivalent width and sSFR is 97.5\%. Comparing these results to the results in Figure~\ref{fig-HI} implies that {\it the CGM properties correlate better with the properties of the atomic gas than those of the current star formation.} 

This result may not be surprising. Warm ionized gas accreted from the CGM has to pass through the atomic phase (traced by \HI 21cm) before condensing into molecular phase, which can then facilitate star formation. The various processes that connect gas inflow to star-formation add intrinsic scatter and therefore, weaken the correlation. For example, the efficiency of the conversion of gas from the atomic to molecular phase, and of molecular gas into stars, may show substantial variation among and within galaxies. The dispersion in the correlation between \Lya equivalent width in the CGM and the SFRs within the galaxies may indicate lags in the multi-step process of accretion of warm gas via the CGM, condensation to atomic gas, conversion to molecular gas, and finally star-formation. 

Nevertheless, we do find a significant difference in the \Lya equivalent width distribution between red and blue galaxies. The blue galaxies show a uniform \Lya strength profile as a function of impact parameter whereas the red galaxies show a much larger dispersion at all impact parameters including small $ \rho\rm /R_{vir}$ (Figure 2). We do find red galaxies with strong \Lya profiles \citep[similar to that found by][]{thom12}, but almost half of the red galaxies either show weak \Lya or even no \Lya at our detection limit. To further probe the relationship between the CGM and star-formation, a paper detailing the analysis of the properties of neutral hydrogen and metal lines from the combined COS-Halos and COS-GASS surveys is in preparation.

\section{Summary \& Implications \label{sec:conclusion}} 

The CGM of galaxies acts as the pathway for baryons and energy to get in and out of the galaxies. The properties of the gas in the CGM not only hold clues to the evolutionary history of the galaxy but also will influence the evolution of the galaxy in the future. After gas from the CGM is accreted onto the galaxy, it is expected to first pass through a phase as atomic hydrogen located primarily in the outer disk.

Therefore, in order to explore the connection between the CGM and the atomic gas reservoir in galactic disks, we have undertaken an observational program with the COS aboard the HST. This program probed the CGM of 45 galaxies out to an impact parameter of $\approx$250~kpc at the rest-frame of the galaxies. The sample covers galaxies in the stellar mass ranges of $\rm 10^{10.1-11.1}~M_{\odot}$, and as such, was designed to probe galaxies spanning the stellar mass range where galaxy the galaxy population transitions from lower mass star-forming galaxies that are rich in cold interstellar gas to more massive quiescent galaxies that contain relatively little cold interstellar gas.

Our sample was chosen from the GASS project, which provided Arecibo \HI spectra, GALEX photometry, SDSS photometry and spectroscopy and all other related data products. For 5 of the galaxies  in the sample that did not have GASS \HI spectra, we obtained \HI spectral measurements with the GBT. 

The COS-GASS sample allowed us to study for the first time the connection between the CGM and the ISM of galaxies. Based on the analysis presented in this paper, we found the following:

\begin{itemize}

\item[1.] We detected \Lya absorption in 92\% (36/39) of the target galaxies for which we have useful data at the expected position of the \Lya transition. Of these detections, 94\%  (34/36) show \Lya absorption within $\pm 250$~\kms of the systemic velocity of their host galaxy.

\item[2.] Combining our sample with the complementary COS Halos sample, we found that the radial distribution of \Lya equivalent widths can be fit as an exponential function with an e-folding scale-length of $\rm R_{vir}$. We defined an 'excess equivalent width' for each \Lya absorption line with respect to the best-fit radial distribution.

\item[3.]  The equivalent widths of the \Lya absorbers do not correlate with orientation of the sightline with respect to the optical major axis of the target galaxy. 

\item[4.] These three results imply that \Lya absorbers are tracing a roughly spherical distribution of gravitationally-bound gas with a characteristic size similar to the virial radius of the dark matter halo. This can indeed be described as a Circum-Galactic Medium. 
 
\item[5.] We find positive correlations (at a confidence of 98.5\%) between the equivalent width of \Lya and $\rm M_{HI}$. An even stronger correlation (at a confidence level of 99.8\%) is present between the excess (impact-parameter-corrected) \Lya equivalent width ($\rm [Log~W ~-~\overline{Log~W}]_{Ly\alpha}$) and the \HI mass fraction ($\rm M_{HI}/M_{\star}$).

\item[6.] Similar correlations were also seen in \Lya spectra stacked as a function of \HI mass and mass fraction. The underlying reason for the observed increase in \Lya absorption strength in the stacks with \HI mass and mass fraction is due to the increase in the covering faction of optically-thick neutral gas. For example, the covering fraction of \Lya absorption is almost 100\% in the sub-sample with \HI mass, M(HI) $> \rm 10^{9.9}~M_{\odot}$. This correlation was observed for sightlines probing both smaller as well as larger impact parameters and is independent of the correlation between the equivalent width of \Lya and impact parameter.

\item[7.] We have used the HI masses to estimate the radii of the HI disks. We then found that the equivalent width of the \Lya line decreases as the ratio of $\rm \rho/R_{HI}$ increases. Sightlines probing the gas within $\sim$ 7~$\rm R_{HI}$ all show strong \Lya absorption features (equivalent widths $ > \rm 0.4 \AA$).

\item[8.] The strength of the \Lya absorption is also correlated with the SFR in their host galaxies. However, we find the correlations between the equivalent width of \Lya and \HI mass and between excess \Lya equivalent width and HI mass fraction to be stronger than the corresponding correlations with SFR and SFR/M$_{\star}$. 

\end{itemize}

These results are consistent with a picture in which the reservoir of cold gas in galaxies is fed by accretion of gas through/from the CGM. In particular, a process that removes or diminishes this CGM reservoir would lead to a subsequent drop in the cold gas content of the galaxy and hence to a diminished star formation rate.

The results above are suggestive of a gas accretion history that is more or less gradual and continuous, at least among the HI-rich galaxies. For a highly episodic process we would expect to find only a weak or no correlation between the CGM and the cold gas content of galaxies. A picture of continuous gas accretion is also consistent with the result that material producing strong (optically-thick) \Lya absorption-lines has a near-unit covering factor in the CGM of the HI-rich star-forming galaxies in our sample. Interestingly, the HI-poor quiescent galaxies show a wider range in the strength of their \Lya absorption features. Many are weak or even non-detections, but we do detect relatively strong absorption feature in others. This might indicate a more sporadic accretion history for them.

It is noteworthy that the properties of the CGM traced by \Lya absorption are more closely correlated with the properties of HI than with the properties of the star formation. This may be explained by the fact that the \HI is mostly in the outer disk and should be more directly related to the effects of accretion from the CGM. In addition, this stronger connection suggests that the material probed with \Lya absorption is primarily gas flowing from the CGM to the galaxy, rather than outflowing gas driven by feedback provided by the ongoing star-formation. This is further supported by the uniformity in the distribution of the \Lya absorbers with respect to the galaxy major/minor axis, and the consistency between the velocity spread of the \Lya absorbers and the 21~cm \HI in the galaxy disk. 

Having established a connection between the cool gas in the CGM and the HI in the galactic disk, it will be important to further probe the interface region between the outer disk and the inner CGM to look for clues as to how and where the accretion/condensation is happening.

\vspace{.5cm}
\acknowledgements 

We thank the HST and GBT staff for support during the observations.
We also thank Cameron Hummels, Jessica Werk, Xavier Prochaska, Hsiao-Wen Chen,  for useful discussions.
This work is based on observations with the NASA/ESA Hubble Space Telescope, which is operated by the Association of Universities for Research in Astronomy, Inc., under NASA contract NAS5-26555. SB and TH were supported by grant HST GO 12603.
BC is the recipient of an Australian Research Council Future Fellowship (FT120100660).

This project also made use of SDSS data. Funding for SDSS-III has been provided by the Alfred P. Sloan Foundation, the Participating Institutions, the National Science Foundation, and the U.S. Department of Energy Office of Science. The SDSS-III web site is http://www.sdss3.org/.

SDSS-III is managed by the Astrophysical Research Consortium for the Participating Institutions of the SDSS-III Collaboration including the University of Arizona, the Brazilian Participation Group, Brookhaven National Laboratory, Carnegie Mellon University, University of Florida, the French Participation Group, the German Participation Group, Harvard University, the Instituto de Astrofisica de Canarias, the Michigan State/Notre Dame/JINA Participation Group, Johns Hopkins University, Lawrence Berkeley National Laboratory, Max Planck Institute for Astrophysics, Max Planck Institute for Extraterrestrial Physics, New Mexico State University, New York University, Ohio State University, Pennsylvania State University, University of Portsmouth, Princeton University, the Spanish Participation Group, University of Tokyo, University of Utah, Vanderbilt University, University of Virginia, University of Washington, and Yale University.

{\it Facilities:}  \facility{Sloan ()} \facility{COS ()} \facility{GBT ()}

\bibliographystyle{apj}	        
\bibliography{myref_bibtex}		

\clearpage
\LongTables 
\begin{landscape}
\begin{deluxetable}{ccccc ccccc   ccccc ccccc}  
\tabletypesize{\scriptsize}
\tablecaption{Description of Galaxy Properties. \label{tbl-galaxy}}
\tablewidth{0pt}
\tablehead{
\colhead{Galaxy} &\colhead{GASS ID} & \colhead{RA} & \colhead{Dec} &\colhead{$\rm z$} &\colhead{$\rm V_{sys}^a$}  & \colhead{$\rm M_{\star} $} & \colhead{$\rm M_{halo}^b$} & \colhead{$\rm R_{vir}$} & \colhead{$\rm M_{HI}$}&\colhead{$\rm \Delta V_{HI}^c$}   &\colhead{$ \rm R_{HI}^d$}  &\colhead{$\rm Source$} &\colhead{sSFR$\rm ^e$}   & \colhead{Color$\rm^f$}    \\
\colhead{}   &\colhead{}   & \colhead{}  & \colhead{}  & \colhead{}  &  \colhead{($\rm km~s^{-1}$)} &\colhead{($\rm Log~M_{\odot}$)} &\colhead{($\rm Log~M_{\odot}$)} & \colhead{(kpc)} & \colhead{($\rm Log~M_{\odot}$)} &  \colhead{($\rm km~s^{-1}$)}  &\colhead{(kpc)}& \colhead{(Survey)} &\colhead{($\rm Log~yr^{-1}$)}  & \colhead{} }
\startdata
J0159+1346  &          3936  &  29.941 & 13.781 & 0.0441 & 13232 & 10.1  &  11.6  &  163  &  9.4  &  -126 $-$ 147
  &  16.2  & GASS  &  -9.5  &  Blue  \\
J0808+0512  &         19852  &  122.068 & 5.216 & 0.0308 & 9227 & 10.8  &  12.5  &  324  &  $<$8.9  &  $-$   &  8.4  & GBT  &  -12.0  &  Red  \\
J0852+0309  &          8096  &  133.229 & 3.152 & 0.0345 & 10351 & 10.3  &  11.8  &  179  &  9.7  &  -195 $-$ 143
  &  22.6  & GASS  &  -10.1  &  Blue  \\
J0908+3234  &         22391  &  137.232 & 32.576 & 0.0490 & 14677 & 10.5  &  12.0  &  212  &  $<$9.3  &  $-$   &  14.0  & GBT  &  $<$-12.3  &  Red  \\
J0914+0836  &         20042  &  138.684 & 8.601 & 0.0468 & 14039 & 10.0  &  11.6  &  156  &  9.0  &  -77 $-$ 108
  &  10.3  & GASS  &  -9.6  &  Blue  \\
J0930+2853  &         32907  &  142.538 & 28.898 & 0.0349 & 10462 & 10.5  &  11.9  &  208  &  9.4  &  -190 $-$ 185
  &  15.0  & GASS  &  -10.7  &  Blue  \\
J0931+2632  &         53269  &  142.817 & 26.550 & 0.0458 & 13718 & 11.0  &  13.0  &  454  &  $<$9.2  &  $-$   &  12.0  & GASS  &  $<$-12.6  &  Red  \\
J0936+3204  &         33214  &  144.101 & 32.079 & 0.0269 & 8073 & 10.3  &  11.8  &  187  &  8.7  &  -31 $-$ 242
  &  6.9  & GASS  &  -11.7  &  Red  \\
J0937+1658  &         55745  &  144.292 & 16.977 & 0.0278 & 8328 & 10.9  &  12.8  &  399  &  8.9  &  -151 $-$ 157
  &  9.1  & GASS  &  -10.3  &  Blue  \\
J0951+3537  &         22822  &  147.937 & 35.622 & 0.0270 & 8091 & 10.6  &  12.1  &  229  &  9.9  &  -145 $-$ 240
  &  31.2  & GASS  &  -10.4  &  Blue  \\
J0958+3204  &         33737  &  149.714 & 32.073 & 0.0270 & 8103 & 10.7  &  12.3  &  269  &  10.0  &  -412 $-$ -146
  &  31.6  & GASS  &  $<$-12.7  &  Red  \\
J1002+3238  &         33777  &  150.711 & 32.645 & 0.0477 & 14300 & 10.1  &  11.6  &  162  &  $<$8.9  &  $-$   &  9.1  & GASS  &  -11.9  &  Red  \\
J1013+0501  &          8634  &  153.352 & 5.025 & 0.0464 & 13916 & 10.1  &  11.6  &  165  &  9.5  &  -187 $-$ 200
  &  17.2  & GASS  &  -10.8  &  Blue  \\
J1032+2112  &         55541  &  158.196 & 21.216 & 0.0429 & 12852 & 10.6  &  12.2  &  245  &  9.9  &  -218 $-$ 200
  &  28.3  & GASS  &  -10.1  &  Blue  \\
J1051+1245  &         23419  &  162.827 & 12.757 & 0.0400 & 11982 & 10.4  &  11.9  &  194  &  10.0  &  -214 $-$ 108
  &  33.6  & ALFALFA  &  -10.0  &  Blue  \\
J1059+0517  &          9109  &  164.811 & 5.292 & 0.0353 & 10591 & 11.1  &  13.3  &  563  &  $<$9.0  &  $-$   &  9.7  & GBT  &  -11.9  &  Red  \\
J1100+1210  &         23457  &  165.048 & 12.171 & 0.0354 & 10606 & 10.1  &  11.6  &  164  &  $<$8.7  &  $-$   &  6.6  & GASS  &  -10.7  &  Blue  \\
J1100+1043  &         23477  &  165.200 & 10.728 & 0.0360 & 10798 & 11.1  &  13.3  &  601  &  9.9  &  $-$   &  29.4  & ALFALFA  &  -11.0  &  Blue  \\
J1115+0241  &          5701  &  168.789 & 2.699 & 0.0442 & 13262 & 10.7  &  12.4  &  284  &  9.8  &  -237 $-$ 227
  &  27.6  & GASS  &  -10.9  &  Blue  \\
J1120+0410  &         12452  &  170.026 & 4.177 & 0.0492 & 14746 & 10.8  &  12.6  &  332  &  $<$9.1  &  $-$   &  10.6  & GASS  &  $<$-12.1  &  Red  \\
J1122+0314  &          5872  &  170.642 & 3.244 & 0.0446 & 13358 & 10.5  &  12.0  &  222  &  $<$9.2  &  $-$   &  12.6  & GBT  &  -12.0  &  Red  \\
J1127+2657  &         48604  &  171.943 & 26.960 & 0.0334 & 10001 & 10.6  &  12.1  &  239  &  9.8  &  -187 $-$ 206
  &  25.6  & GASS  &  -11.0  &  Blue  \\
J1131+1553  &         29898  &  172.954 & 15.897 & 0.0364 & 10924 & 10.2  &  11.7  &  169  &  $<$9.0  &  $-$   &  10.1  & GBT  &  -12.0  &  Red  \\
J1132+1329  &         29871  &  173.052 & 13.492 & 0.0342 & 10267 & 10.2  &  11.7  &  168  &  9.8  &  -256 $-$ 251
  &  25.0  & ALFALFA  &  -9.7  &  Blue  \\
J1142+3013  &         48994  &  175.575 & 30.230 & 0.0322 & 9665 & 10.7  &  12.4  &  287  &  10.3  &  -196 $-$ 227
  &  49.4  & GASS  &  -10.4  &  Blue  \\
J1155+2921  &         49433  &  178.903 & 29.351 & 0.0458 & 13718 & 10.5  &  11.9  &  205  &  9.3  &  -148 $-$ 162
  &  13.6  & GASS  &  -10.3  &  Blue  \\
J1241+2847  &         50550  &  190.367 & 28.791 & 0.0350 & 10498 & 10.3  &  11.8  &  180  &  9.4  &  -125 $-$ 177
  &  16.4  & GASS  &  -10.0  &  Blue  \\
J1251+0551  &         13074  &  192.894 & 5.864 & 0.0486 & 14572 & 10.9  &  12.6  &  353  &  10.2  &  -257 $-$ 291
  &  41.5  & ALFALFA  &  -10.4  &  Blue  \\
J1305+0359  &         13159  &  196.356 & 3.992 & 0.0437 & 13094 & 10.4  &  11.8  &  191  &  9.4  &  -127 $-$ 262
  &  15.6  & GASS  &  -10.8  &  Blue  \\
J1315+1525  &         26936  &  198.855 & 15.423 & 0.0266 & 7959 & 10.7  &  12.4  &  294  &  $<$8.9  &  $-$   &  9.1  & GASS  &  $<$-12.3  &  Red  \\
J1317+2629  &         51025  &  199.440 & 26.486 & 0.0450 & 13490 & 10.3  &  11.7  &  176  &  9.8  &  -133 $-$ 101
  &  27.6  & ALFALFA  &  -10.4  &  Blue  \\
J1325+2714  &         51161  &  201.345 & 27.249 & 0.0345 & 10336 & 10.1  &  11.7  &  166  &  9.5  &  -123 $-$ 183
  &  18.1  & GASS  &  -9.8  &  Blue  \\
J1348+2453  &         38018  &  207.142 & 24.891 & 0.0297 & 8903 & 10.1  &  11.6  &  160  &  9.5  &  -173 $-$ 173
  &  17.0  & GASS  &  -10.5  &  Blue  \\
J1354+2433  &         44856  &  208.546 & 24.556 & 0.0286 & 8583 & 10.1  &  11.6  &  158  &  9.3  &  -160 $-$ 187
  &  13.3  & GASS  &  -11.8  &  Red  \\
J1404+3357  &         31172  &  211.122 & 33.953 & 0.0264 & 7920 & 10.3  &  11.8  &  181  &  9.5  &  -193 $-$ 232
  &  17.2  & CHA  &  $<$-12.3  &  Red  \\
J1406+0154  &          7121  &  211.678 & 1.915 & 0.0472 & 14153 & 10.2  &  11.7  &  175  &  8.9  &  -188 $-$ 144
  &  8.7  & GASS  &  -11.8  &  Red  \\
J1427+2629  &         45940  &  216.954 & 26.484 & 0.0325 & 9734 & 10.4  &  11.9  &  200  &  $<$8.7  &  $-$   &  6.7  & GASS  &  -12.0  &  Red  \\
J1430+0323  &          9615  &  217.508 & 3.398 & 0.0333 & 9983 & 10.2  &  11.7  &  166  &  9.4  &  -133 $-$ 154
  &  16.2  & GASS  &  -11.1  &  Red  \\
J1431+2440  &         38198  &  217.894 & 24.682 & 0.0378 & 11338 & 10.7  &  12.2  &  256  &  9.7  &  -211 $-$ 247
  &  23.2  & GASS  &  $<$-12.7  &  Red  \\
J1454+3050  &         42191  &  223.516 & 30.846 & 0.0320 & 9584 & 10.1  &  11.6  &  164  &  8.8  &  -173 $-$ 97
  &  7.8  & GASS  &  -9.8  &  Blue  \\
J1502+0649  &         41743  &  225.517 & 6.823 & 0.0462 & 13859 & 10.5  &  11.9  &  204  &  9.6  &  -166 $-$ 186
  &  20.5  & GASS  &  -10.2  &  Blue  \\
J1509+0704  &         41869  &  227.340 & 7.078 & 0.0414 & 12414 & 10.1  &  11.7  &  166  &  9.3  &  -133 $-$ 176
  &  14.0  & GASS  &  -9.6  &  Blue  \\
J1515+0701  &         42025  &  228.781 & 7.021 & 0.0367 & 11005 & 10.9  &  12.7  &  368  &  9.5  &  -206 $-$ 205
  &  17.5  & GASS  &  -11.9  &  Red  \\
J1541+2813  &         28365  &  235.344 & 28.230 & 0.0321 & 9620 & 10.4  &  11.8  &  189  &  10.1  &  -63 $-$ 103
  &  36.2  & CHA  &  -9.6  &  Blue  \\
J1544+2740  &         28317  &  236.034 & 27.673 & 0.0316 & 9482 & 10.1  &  11.6  &  160  &  9.1  &  -19 $-$ 191
  &  10.9  & GASS  &  $<$-12.1  &  Red
\enddata
\tablenotetext{a}{Systemic velocity, $\rm V_{sys}$, is the velocity corresponding to the redshift of the galaxy i.e. $\rm V_{sys}= ~c~z$, where $c$ is the speed of light in vacuum.}
\tablenotetext{b}{Using prescription from Behroozi et al. 2010  (Eq. 21).}
\tablenotetext{c}{The full width of the \textsc{Hi} 21cm line in the rest-frame of the galaxy. The systemic velocity was subtacted from the observed \textsc{Hi}  velocity to obtaine the rest-frame velocity.}
\tablenotetext{d}{Using emperical relationship relating M$\rm _{HI}$ and R$\rm _{HI}$ as prescribed by \citet{swaters02}.}
\tablenotetext{e}{Values are from \citet{ schiminovich10,catinella10}. Values for sSFR $ \rm < 10^{-12}~ yr^{-1}$ are regarded as upper limits \citep[discussed in ][]{schiminovich10}.}
\tablenotetext{f}{Galaxies with sSFR $>\rm 10^{-11}~yr^{-1}$ are defined as blue galaxies.  Galaxies with sSFR below this value are identified as red galaxies.}
\end{deluxetable}
\clearpage
\end{landscape}

\clearpage
\LongTables 
\begin{landscape}
\begin{deluxetable}{ccccc ccccc   ccccc ccccc c}  
\tabletypesize{\scriptsize}
\tablecaption{Description of QSO Sightlines. \label{tbl-sightlines}}
\tablewidth{0pt}
\tablehead{
\colhead{QSO} & \colhead{RA$_{\rm QSO}$} & \colhead{Dec$_{\rm QSO}$} & \colhead{z$_{\rm QSO}$} & \colhead{Galaxy} & \colhead{$\rho$} & \colhead{$\rm \rho/R_{vir}$} & \colhead{$\rm \Theta^a$}   & \colhead{$\rm W_{Ly\alpha}$} & \colhead{$\rm W_{Ly\alpha}/ \bar{~W}_{Ly\alpha}$} & \colhead{$\rm v_{\rm Ly\alpha}^b$} & \colhead{$\rm \Delta v_{Ly\alpha}^c$}    & \colhead{$\rm v_{HI}^d$} & \colhead{$\rm v_{esc,\rho}^e$}   & \colhead{$\rm v_{esc,Rvir}^e$} \\
\colhead{}    & \colhead{}    &   \colhead{}             & \colhead{}                & \colhead{}              & \colhead{(kpc)}  & \colhead{}                   & \colhead{}                 & \colhead{($\rm \AA$)}        & \colhead{}               & \colhead{($\rm km~s^{-1}$)}   & \colhead{($\rm km~s^{-1}$)}  &  \colhead{($\rm km~s^{-1}$)}& \colhead{($\rm km~s^{-1}$)}& \colhead{($\rm km~s^{-1}$)} }
\startdata
J0159+1345  &  29.971  &  13.765  &  0.504  &  J0159+1346  &  102  &  0.6  &  64 & 1.501$\pm$0.023  &  2.975 & 93,358  &  -150 $-$ 380
 &          147 & 178 & 150  \\
J0808+0514  &  122.162  &  5.244  &  0.361  &  J0808+0512  &  215  &  0.7  &  7 &  $-$  &   $-$  & $-$  &  $-$  &  $-$  & 348 & 300  \\
J0852+0313  &  133.247  &  3.222  &  0.297  &  J0852+0309  &  178  &  1.0  &  67 & 0.113$\pm$0.015  &  0.322 & 50  &  0 $-$ 110
 &          143 & 165 & 165  \\
J0909+3236  &  137.276  &  32.608  &  0.809  &  J0908+3234  &  170  &  0.8  &  21 & 0.090$\pm$0.014  &  0.213 & 101  &  50 $-$ 200
 &  $-$  & 212 & 195  \\
J0914+0837  &  138.632  &  8.629  &  0.649  &  J0914+0836  &  189  &  1.2  &  69 & 0.104$\pm$0.020  &  0.370 & 30  &  -50 $-$ 80
 &          108 & 133 & 143  \\
J0930+2848  &  142.508  &  28.816  &  0.487  &  J0930+2853  &  214  &  1.0  &  42 &  $<$0.126  &  $<$0.372 & $-$  &  $-$  &  $-$  & 189 & 191  \\
J0931+2628  &  142.820  &  26.480  &  0.778  &  J0931+2632  &  226  &  0.5  &  77 & 0.114$\pm$0.013  &  0.199 & 200  &  150 $-$ 250
 &  $-$  & 541 & 423  \\
J0936+3207  &  144.016  &  32.119  &  1.150  &  J0936+3204  &  160  &  0.9  &  0$^f$ &  $<$0.113  &  $<$0.280 & $-$  &  $-$  &  $-$  & 183 & 172  \\
J0937+1700  &  144.279  &  17.006  &  0.506  &  J0937+1658  &  63  &  0.2  &  64 & 0.135$\pm$0.021  &  0.168 & -204  &  -300 $-$ -150
 &         -151 & 669 & 370  \\
J0951+3542  &  147.850  &  35.714  &  0.398  &  J0951+3537  &  226  &  1.0  &  3 & 0.839$\pm$0.014  &  2.382 & 64  &  -90 $-$ 200
 &          240 & 212 & 211  \\
J0959+3203  &  149.812  &  32.066  &  0.564  &  J0958+3204  &  162  &  0.6  &  0$^f$ & 0.420$\pm$0.016  &  0.815 & -238,-144  &  -320 $-$ -100
 &         -412 & 298 & 249  \\
J1002+3240  &  150.727  &  32.678  &  0.829  &  J1002+3238  &  119  &  0.7  &  40 &  $<$0.063  &  $<$0.140 & $-$  &  $-$  &  $-$  & 166 & 149  \\
J1013+0500  &  153.325  &  5.009  &  0.266  &  J1013+0501  &  102  &  0.6  &  0$^f$ & 0.445$\pm$0.024  &  0.875 & -72  &  -180 $-$ 50
 &         -187 & 179 & 151  \\
J1033+2112  &  158.270  &  21.204  &  0.315  &  J1032+2112  &  214  &  0.9  &  45 & 0.437$\pm$0.033  &  1.104 & -1  &  -100 $-$ 95
 &         -218 & 238 & 227  \\
J1051+1247  &  162.857  &  12.796  &  1.281  &  J1051+1245  &  140  &  0.7  &  73 & 0.781$\pm$0.022  &  1.706 & -37  &  -180 $-$ 140
 &         -214 & 201 & 178  \\
J1059+0519  &  164.795  &  5.327  &  0.754  &  J1059+0517  &  95  &  0.2  &  30 & 0.270$\pm$0.022  &  0.340 & 6  &  -100 $-$ 100
 &  $-$  & 932 & 525  \\
J1059+1211  &  164.984  &  12.198  &  0.993  &  J1100+1210  &  171  &  1.0  &  61 & 0.225$\pm$0.014  &  0.676 & -61  &  -150 $-$ 0
 &  $-$  & 148 & 151  \\
J1100+1046  &  165.199  &  10.770  &  0.422  &  J1100+1043  &  108  &  0.2  &  0$^f$ &  $-$  &   $-$  & $-$  &  $-$  &  $-$  & 980 & 562  \\
J1115+0237  &  168.782  &  2.633  &  0.567  &  J1115+0241  &  209  &  0.7  &  83 & 0.195$\pm$0.020  &  0.430 & 2  &  -100 $-$ 100
 &          227 & 293 & 263  \\
J1120+0413  &  170.021  &  4.223  &  0.545  &  J1120+0410  &  162  &  0.5  &  78 & 0.830$\pm$0.018  &  1.434 & 201,370  &  50 $-$ 390
 &  $-$  & 396 & 308  \\
J1122+0318  &  170.601  &  3.301  &  0.475  &  J1122+0314  &  221  &  1.0  &  0$^f$ & 0.110$\pm$0.018  &  0.315 & 67  &  0 $-$ 150
 &  $-$  & 205 & 204  \\
J1127+2654  &  171.902  &  26.914  &  0.379  &  J1127+2657  &  140  &  0.6  &  26 & 0.705$\pm$0.021  &  1.342 & -28,-178  &  -250 $-$ 100
 &         -187 & 267 & 220  \\
J1131+1556  &  172.905  &  15.946  &  0.183  &  J1131+1553  &  176  &  1.0  &  0$^f$ &  $-$  &   $-$  & $-$  &  $-$  &  $-$  & 153 & 155  \\
J1132+1335  &  173.044  &  13.586  &  0.201  &  J1132+1329  &  230  &  1.4  &  5 & 0.319$\pm$0.017  &  1.318 & 57  &  -40 $-$ 130
 &          251 & 138 & 155  \\
J1142+3016  &  175.551  &  30.270  &  0.481  &  J1142+3013  &  104  &  0.4  &  50 & 0.886$\pm$0.023  &  1.351 & 3  &  -200 $-$ 170
 &          227 & 376 & 265  \\
J1155+2922  &  178.970  &  29.377  &  0.520  &  J1155+2921  &  208  &  1.0  &  1 & 0.742$\pm$0.023  &  2.156 & 72,-150  &  -200 $-$ 230
 &          162 & 188 & 189  \\
J1241+2852  &  190.374  &  28.870  &  0.589  &  J1241+2847  &  198  &  1.1  &  40 & 0.211$\pm$0.020  &  0.667 & 33  &  -120 $-$ 170
 &          177 & 160 & 166  \\
J1251+0554  &  192.853  &  5.906  &  1.377  &  J1251+0551  &  200  &  0.6  &  57 & 0.409$\pm$0.021  &  0.765 & 80  &  -20 $-$ 180
 &          291 & 401 & 328  \\
J1305+0357  &  196.351  &  3.959  &  0.545  &  J1305+0359  &  103  &  0.5  &  11 & 0.821$\pm$0.016  &  1.494 & 67,-43  &  -160 $-$ 180
 &          262 & 218 & 175  \\
J1315+1525  &  198.938  &  15.432  &  0.448  &  J1315+1525  &  155  &  0.5  &  17 & 0.405$\pm$0.018  &  0.728 & 64  &  -50 $-$ 170
 &  $-$  & 341 & 272  \\
J1318+2628  &  199.508  &  26.475  &  1.234  &  J1317+2629  &  198  &  1.1  &  86 & 0.184$\pm$0.032  &  0.597 & 0  &  -120 $-$ 120
 &         -133 & 155 & 162  \\
J1325+2717  &  201.266  &  27.289  &  0.522  &  J1325+2714  &  199  &  1.2  &  55 &  $-$  &   $-$  & $-$  &  $-$  &  $-$  & 143 & 153  \\
J1348+2456  &  207.093  &  24.947  &  0.293  &  J1348+2453  &  153  &  1.0  &  81 & 0.474$\pm$0.035  &  1.301 & -97  &  -230 $-$ 0
 &         -173 & 150 & 147  \\
J1354+2430  &  208.604  &  24.502  &  1.878  &  J1354+2433  &  155  &  1.0  &  78 & 0.545$\pm$0.034  &  1.537 & -97,-5  &  -170 $-$ 50
 &         -160 & 147 & 146  \\
J1404+3353  &  211.118  &  33.895  &  0.549  &  J1404+3357  &  111  &  0.6  &  57 & 0.749$\pm$0.027  &  1.469 & -26  &  -150 $-$ 150
 &         -193 & 198 & 166  \\
J1406+0157  &  211.732  &  1.954  &  0.427  &  J1406+0154  &  222  &  1.3  &  67 &  $-$  &   $-$  & $-$  &  $-$  &  $-$  & 147 & 161  \\
J1427+2632  &  216.898  &  26.537  &  0.364  &  J1427+2629  &  170  &  0.9  &  0$^f$ &  $-$  &   $-$  & $-$  &  $-$  &  $-$  & 195 & 184  \\
J1429+0321  &  217.420  &  3.357  &  0.253  &  J1430+0323  &  231  &  1.4  &  0$^f$ & 0.807$\pm$0.027  &  3.414 & -39,109  &  -150 $-$ 250
 &         -133 & 135 & 153  \\
J1431+2442  &  217.858  &  24.706  &  0.407  &  J1431+2440  &  110  &  0.4  &  18 & 0.569$\pm$0.015  &  0.929 & 73,156  &  0 $-$ 220
 &          247 & 317 & 236  \\
J1454+3046  &  223.601  &  30.783  &  0.465  &  J1454+3050  &  223  &  1.4  &  37 & 0.472$\pm$0.035  &  1.941 & 57  &  -50 $-$ 160
 &           97 & 134 & 151  \\
J1502+0645  &  225.517  &  6.754  &  0.288  &  J1502+0649  &  224  &  1.1  &  80 & 0.438$\pm$0.013  &  1.384 & 12,-55  &  -150 $-$ 110
 &          186 & 182 & 188  \\
J1509+0702  &  227.368  &  7.043  &  0.418  &  J1509+0704  &  130  &  0.8  &  62 & 0.956$\pm$0.022  &  2.220 & 49,-18,-215  &  -275 $-$ 130
 &          176 & 167 & 153  \\
J1515+0657  &  228.781  &  6.952  &  0.268  &  J1515+0701  &  180  &  0.5  &  14 & 0.270$\pm$0.023  &  0.468 & -367  &  -500 $-$ -300
 &         -206 & 439 & 342  \\
J1541+2817  &  235.340  &  28.285  &  0.376  &  J1541+2813  &  128  &  0.7  &  0$^f$ & 0.864$\pm$0.011  &  1.797 & -363  &  -520 $-$ -250
 &          -63 & 200 & 174  \\
J1544+2743  &  236.114  &  27.723  &  0.163  &  J1544+2740  &  196  &  1.2  &  55 & 0.191$\pm$0.022  &  0.687 & 126  &  50 $-$ 210
 &          191 & 136 & 147
\enddata
\tablenotetext{a}{Orientation of the QSO sightlines with respect to the disk of the galaxies. The values are based on SDSS r-band photometric measurements.}
\tablenotetext{b}{Centroid of the multiple components of the \Lya absorption feature as estimated via Voigt profile fitting in the order of the strength of the component.}
\tablenotetext{c}{Full width of the absorption feature in the rest-frame of the galaxy.}
\tablenotetext{d}{Maximum velocity of the \textsc{Hi} profile in the rest-frame of the galaxy. This value was measured on the velocity side where the centroid of the Lyman~$\alpha$ absorber was detected.}
\tablenotetext{e}{Line of sight escape velocity at the impact parameter probed by the QSO sightline assuming a NFW profile for the galaxy's dark matter distribution.}
\tablenotetext{f}{Face-on galaxies.}
\end{deluxetable}
\clearpage
\end{landscape}

\end{document}